\documentclass[
 reprint,
 superscriptaddress,
 nofootinbib,
 amsmath,amssymb,
 aps,
 prl
]{revtex4-2}

\usepackage{lipsum}  % Added by Peppe
\usepackage{csquotes} % Added by Peppe
\usepackage[caption = false]{subfig} % Added by Peppe

\usepackage{graphicx}% Include figure files
\usepackage{dcolumn}% Align table columns on decimal point
\usepackage{bm}% bold math
\usepackage[normalem]{ulem}
\usepackage{xcolor}
\begin{document}

\preprint{APS/123-QED}

\title{Spatio-Temporal Energy Cascade in Three-Dimensional Magnetohydrodynamic Turbulence}

\author{Giuseppe Arrò}
\affiliation{Los Alamos National Laboratory, Los Alamos, NM 87545, USA}

\author{Hui Li}
\affiliation{Los Alamos National Laboratory, Los Alamos, NM 87545, USA}

\author{William H. Matthaeus}
\affiliation{Department of Physics and Astronomy, University of Delaware, Newark, DE 19716, USA}

%\date{\today}

\begin{abstract}

\vspace{1pt}

We present a new scale decomposition method to investigate turbulence in wavenumber-frequency space. Using 3D magnetohydrodynamic turbulence simulations, we show that magnetic fluctuations with time scales longer than the nonlinear time exhibit an inverse cascade toward even smaller frequencies. Low frequency magnetic fluctuations support turbulence, acting as an energy reservoir that is converted into plasma kinetic energy, the latter cascading toward large wavenumbers and frequencies, where it is dissipated. Our results shed new light on the spatio-temporal properties of turbulence, potentially explaining the origin and role of low frequency turbulent fluctuations in the solar wind.

\end{abstract}

%\keywords{turbulence}

\maketitle

Turbulence is ubiquitous in space and astrophysical plasmas, including the solar wind (SW) \citep{bruno2013solar}, the interstellar medium \citep{armstrong1995electron}, accretion disks of compact objects \citep{balbus1998instability}, and galaxy clusters \citep{zhuravleva2014turbulent}. The signature of turbulence is the generation of fluctuations with broadband wavenumber and frequency spectra. Turbulent plasmas are typically subject to strong background magnetic fields, inducing a wavenumber anisotropy of the kind $k_{\perp}\!>\!k_{\parallel}$ \citep{shebalin1983anisotropy,goldreich1995toward,cho2000anisotropy,horbury2008anisotropic,wicks2010power,cerri2019kinetic,oughton2020critical}, with $k_{\perp}$ and $k_{\parallel}$ representing wavenumbers perpendicular and parallel to the magnetic field. Numerical \citep{parashar2010kinetic,markovskii2020four,papini2021spacetime,fu2022nature,gan2022existence} and observational \citep{narita2010magnetic,zhao2023observations} SW studies have revealed that plasma turbulence has nontrivial spatio-temporal properties, with a tendency for magnetic field, velocity and density fluctuations to be concentrated at low frequencies $\omega$ and small $k_{\parallel}$. Low $(k_{\parallel},\,\omega)$ modes are often interpreted as quasi-2D structures, like flux ropes and vortices \citep{alexandrova2008solar,lion2016coherent,zank2021turbulence,arro2023generation,arro2024large,zank2024characterization}, since their energy distribution in $(k_{\perp},\,k_{\parallel},\,\omega)$ space does not follow the dispersion relation of waves. 

The origin of low $(k_{\parallel},\,\omega)$ turbulent fluctuations has been debated for decades and is still not well understood. Several theoretical explanations for the formation of $(\omega\!\sim\!0,\,k_{\parallel}\!\sim\!0)$ modes have been proposed, including quasi-2D models of nearly incompressible turbulence \citep{matthaeus1990evidence,zank1992waves,zank2011transport}, turbulence driven by counterpropagating Alfvén waves \citep{ng1996interaction,galtier2000weak}, nonlinear frequency broadening of magnetohydrodynamic (MHD) modes \citep{dmitruk2009waves,yuen2023temporal}, and inverse turbulent cascades driven by the conservation of MHD and reduced-MHD invariants \citep{dmitruk2007low,dmitruk2011emergence}. In addition to understanding their origin, the presence of low $(k_{\parallel},\,\omega)$ modes also raises the question of whether such fluctuations play a role in driving the turbulent cascade, or if they are dynamically irrelevant once generated. Despite the longstanding efforts in addressing these questions, a unifying description for the origin of low $(k_{\parallel},\,\omega)$ fluctuations and their role in turbulence is still missing. 

Understanding the spatio-temporal properties of turbulence has fundamental implications for several space and astrophysical problems. These include the solar corona heating \cite{matthaeus1999coronal,dmitruk2002coronal}, the SW acceleration and expansion \citep{cranmer2007self,cranmer2015role}, the emergence of \enquote{$1/f$} frequency spectra in the SW \citep{matthaeus1986low,matthaeus2007density}, the acceleration and transport of cosmic rays \citep{giacalone1999transport,yan2002scattering,yan2004cosmic}, angular momentum transport in accretion disks \citep{klahr2003turbulence,sano2004angular}, the dynamics of molecular clouds and star formation \citep{larson1981turbulence,schneider2011link}, and the turbulent dynamo \citep{schekochihin2004simulations,ponty2004simulation,beresnyak2012universal}.

In this Letter, we present a new framework to study how turbulence reorganizes energy among fluctuations with different wavenumbers and frequencies. We use this method to investigate the origin of low frequency SW turbulent fluctuations, and their role in the turbulent cascade. Our approach employs the coarse graining (CG) technique, frequently used to study hydrodynamic \citep{eyink2009localness,aluie2012conservative,mondal2024spatio}, MHD \citep{aluie2010scale,aluie2013scale,yang2016energy}, and plasma turbulence \citep{yang2017energy,yang2017compressibility,matthaeus2020pathways,arro2022spectral}. The CG method consists in low-pass filtering the equations of motion of the system, cutting-off small scales. The global energy balance obtained from filtered equations gives a set of quantities describing large scale energy transfers, plus some cascade terms, representing energy exchanges between large and small scales. We apply this method to MHD equations
\begin{equation}
\begin{gathered}
\partial_t \rho + \nabla \!\cdot\! \bigl( \rho \textbf{u} \bigr) \!= 0,
\\[5pt]
\partial_t \bigl( \rho \textbf{u} \bigr)\! + \!\nabla \!\cdot\! \bigl( \rho \textbf{u} \textbf{u} \bigr)\! = -\!\nabla P + \textbf{J} \times \textbf{B} + \!\nabla \!\cdot\! \boldsymbol{\Pi} + \textbf{F}_u,
\\[5pt]
\partial_t \textbf{B} = \!\nabla \!\times\! \bigl( \textbf{u} \times \textbf{B} \!-\!\eta\,\textbf{J} \bigr)\! + \textbf{F}_{_B}, 
\end{gathered}
\label{FMHD}
\end{equation}
where $\rho$, $\textbf{u}$ and $P$ are the plasma density, velocity and pressure. $\textbf{B}$ is the magnetic field, $\textbf{J}\!=\!\nabla\!\times\!\textbf{B}$, $\eta$ is the magnetic diffusivity. $\boldsymbol{\Pi} \!=\! \rho\,\nu \bigl[ \nabla\textbf{u} + \!\nabla\textbf{u}^T\! - \!\bigl( 2/3 \bigr) \bigl(\nabla\!\cdot\!\textbf{u}\bigr) \textbf{I} \bigr]$ is the viscous stress tensor, where $\nu$ is the viscosity, $\textbf{I}$ is the identity matrix, and $T$ indicates the transpose operation. $\textbf{F}_u$ and $\textbf{F}_{_B}$ are large scale turbulent forcing terms. 

The CG method typically involves only spatial scales, but we extend the technique by introducing the spatio-temporal low-pass filter
\begin{equation}
\overline{q}\bigl(\textbf{x},\,t,\,\textbf{k},\,\tau\bigr)\!=\! \sum\limits_{\textbf{k}^{\prime}<\textbf{k}}\int dt^{\prime} \,Q\bigl(\textbf{k}^{\prime},\,t^{\prime}\bigr)\,e^{i\,\textbf{k}^{\prime}\cdot\textbf{x}}\,G_{\tau}\bigl(t\!-\!t^{\prime}\bigr),
\end{equation}
where $q(\textbf{x},\,t)$ is a generic quantity with spatial Fourier transform $Q(\textbf{k},\,t)$, and $G_{\tau}$ is a boxcar function with width $\tau$. $\overline{q}$ contains wavenumbers $<\textbf{k}$ and time scales $>\tau$. The corresponding density-weighted filter is $\widehat{q}\!=\!\overline{\rho\,q}/\overline{\rho}$ \citep{favre1969problems}. By applying this filter to Eqs.~(\ref{FMHD}), and calculating the global energy balance, we obtain:
\begin{equation}
\begin{gathered}
\partial_t \biggl< \frac{1}{2}\,\overline{\rho}\,\widehat{u}^2 \biggr> = -W_{_P} -\Pi_u +W_{e.m.} -D_u +I_u,
\\[5pt]
\partial_t \biggl< \frac{1}{2}\,\overline{B}^2 \biggr> = -\Pi_{_B} -W_{e.m.} -D_{_B} +I_{_B},
\end{gathered}
\label{CGMHD}
\end{equation}
where $\langle \cdot \rangle$ indicates the spatial average over the system size, and we assumed no transport across the system boundaries, causing all terms in the form of a divergence to vanish when averaged (as in periodic systems). A detailed derivation of Eqs.~(\ref{CGMHD}) is provided in Supplemental Material. The energy transfer channels (ETCs) in Eqs.~(\ref{CGMHD}) are
\begin{equation}
\begin{gathered}
I_{_B}\!=\!\bigl< \overline{\textbf{F}}_{_B} \!\cdot \overline{\textbf{B}} \bigr>,
\quad
I_u\!=\!\bigl< \overline{\textbf{F}}_u \!\cdot \widehat{\textbf{u}} \bigr>,
\\[7pt]
W_{e.m.}\!=\!\bigl< -\bigl( \widehat{\textbf{u}}\times\overline{\textbf{B}} \bigr) \!\cdot \overline{\textbf{J}} \bigr>, 
\quad
\Pi_{_B}\!=\!\bigl< \boldsymbol{\tau}_{_E} \!\cdot \overline{\textbf{J}} \bigr>,
\\[7pt]
\Pi_u\!=\! \Pi_u^{S} + \Pi_u^{L} \!=\! \bigl< - \overline{\rho}\, \boldsymbol{\tau}_u \!:\! \nabla\widehat{\textbf{u}} \bigr> + \bigl< - \boldsymbol{\tau}_{_B} \!\cdot \widehat{\textbf{u}} \bigr>, 
\\[7pt]
W_{_P}\!=\!\bigl< -\overline{P}\, \nabla\!\cdot\!\widehat{\textbf{u}} \bigr>,
\quad
D_{_B}\!=\!\bigl< \eta\, \overline{\textbf{J}}^2 \bigr>,
\quad
D_u\!=\!\bigl< \overline{\boldsymbol{\Pi}} : \!\nabla\widehat{\textbf{u}} \bigr>,
\end{gathered}
\label{ETC}
\end{equation}
with $I_{_B}$ and $I_u$ representing the magnetic and kinetic energy injection rates, $W_{e.m.}$ and $W_{_P}$ are the electromagnetic (e.m.) and pressure works, $D_{_B}$ and $D_u$ are the magnetic and kinetic energy dissipation rates, all representing energy exchanges at scales $<\textbf{k}$ and $>\tau$. $\Pi_{_B}$ and $\Pi_u$ are the magnetic and kinetic energy cascade rates, quantifying energy transfers from scales $<\textbf{k}$ and $>\tau$, to scales $>\textbf{k}$ and $<\tau$. $\boldsymbol{\tau}_u\!=\!(\widehat{\textbf{u}\textbf{u}}-\widehat{\textbf{u}}\widehat{\textbf{u}})$ is the subscale stress tensor \citep{eyink2006multi}, while $\boldsymbol{\tau}_{_E}\!=\!-(\overline{\textbf{u}\times\textbf{B}}-\widehat{\textbf{u}}\times\overline{\textbf{B}})$ and $\boldsymbol{\tau}_{_B}\!=\!(\overline{\textbf{J}\times\textbf{B}}-\overline{\textbf{J}}\times\overline{\textbf{B}})$ are the subscale electric field and Lorentz force. Subscale terms couple small scales to large scales. Hence, $\Pi_{_B}$ represents the interaction between small scale electric fields $\boldsymbol{\tau}_{_E}$ and large scale currents $\overline{\textbf{J}}$. $\Pi_u$ includes two contributions: $\Pi_u^S$ quantifies the interaction between small scale stresses $\boldsymbol{\tau}_u$ and the large scale strain tensor $\nabla\widehat{\textbf{u}}$; $\Pi_u^L$ couples the small scale Lorentz force $\boldsymbol{\tau}_{_B}$ to large scale velocities $\widehat{\textbf{u}}$. 

We apply our CG method to a 3D simulation of MHD turbulence, realized with \textit{Athena++} \citep{Stone2020}, implementing Eqs.~(\ref{FMHD}). We consider a uniform periodic grid with $256\!\times\!512^2$ points, and size $L_z\!=\!3L_y\!=\!3L_x\!=\!6\,\pi$ (in arbitrary units $L_0$). The plasma has zero initial velocity, homogeneous density $\rho_0$ and guide field $\textbf{B}_0\!=\!B_0\widehat{\textbf{z}}$. Pressure is isothermal, with plasma beta $\beta\!=\!2\,c^2_{_S}/c^2_{_A}\!=\!0.5$, where $c_{_S}$ and $c_{_A}\!=\!B_0/\sqrt{\rho_0}$ are the sound and Alfvén speeds. Turbulence is driven by $\textbf{F}_u$ and $\textbf{F}_{_B}$, consisting of sinusoidal perturbations with wavenumbers $1\!\leqslant\!k_{\parallel}/k_{\parallel,0}\!\leqslant\!3$ and $1\!\leqslant\!k_{\perp}/k_{\perp,0}\!\leqslant\!4$ (where $k_{\parallel,0}\!=\!2\pi/L_z$ and $k_{\perp,0}\!=\!2\pi/L_x$). Each perturbation evolves in time following the Langevin antenna (LA) scheme \citep{tenbarge2014oscillating}, with driving frequency $\omega_0\!=\!0.8\,\tau_{_A}^{-1}$ and decorrelation rate $\gamma_0\!=\!-0.7\,\tau_{_A}^{-1}$ (where $\tau_{_A}\!=\!L_0/c_{_A}$). $\textbf{F}_u$ and $\textbf{F}_{_B}$ are solenoidal and perpendicular to $\textbf{B}_0$, driving incompressible Alfvénic perturbations with cross helicity $\sigma_{_C}\!\simeq\!0$. $\textbf{F}_u$ and $\textbf{F}_{_B}$ are set to induce magnetic and velocity fluctuations with root mean square (rms) amplitudes $\delta B_{rms}/B_0\!\simeq\!0.18$ and $\delta u_{rms}/c_{_A}\!\simeq\!0.17$. Our setup produces turbulence with typical near-Earth SW parameters \citep{bandyopadhyay2020situ}. We set $\eta\!=\!\nu\!=\!2.5\cdot10^{-4}$ (in $L_0^2/\tau_{_A}$ units).

%%%%%%%%%%%%%%%%%%%%%%%%%%%%%%%%%%%%%%%%%%%%%%%%%%
\begin{figure}[t]
\centering
\includegraphics[width=0.98\linewidth]{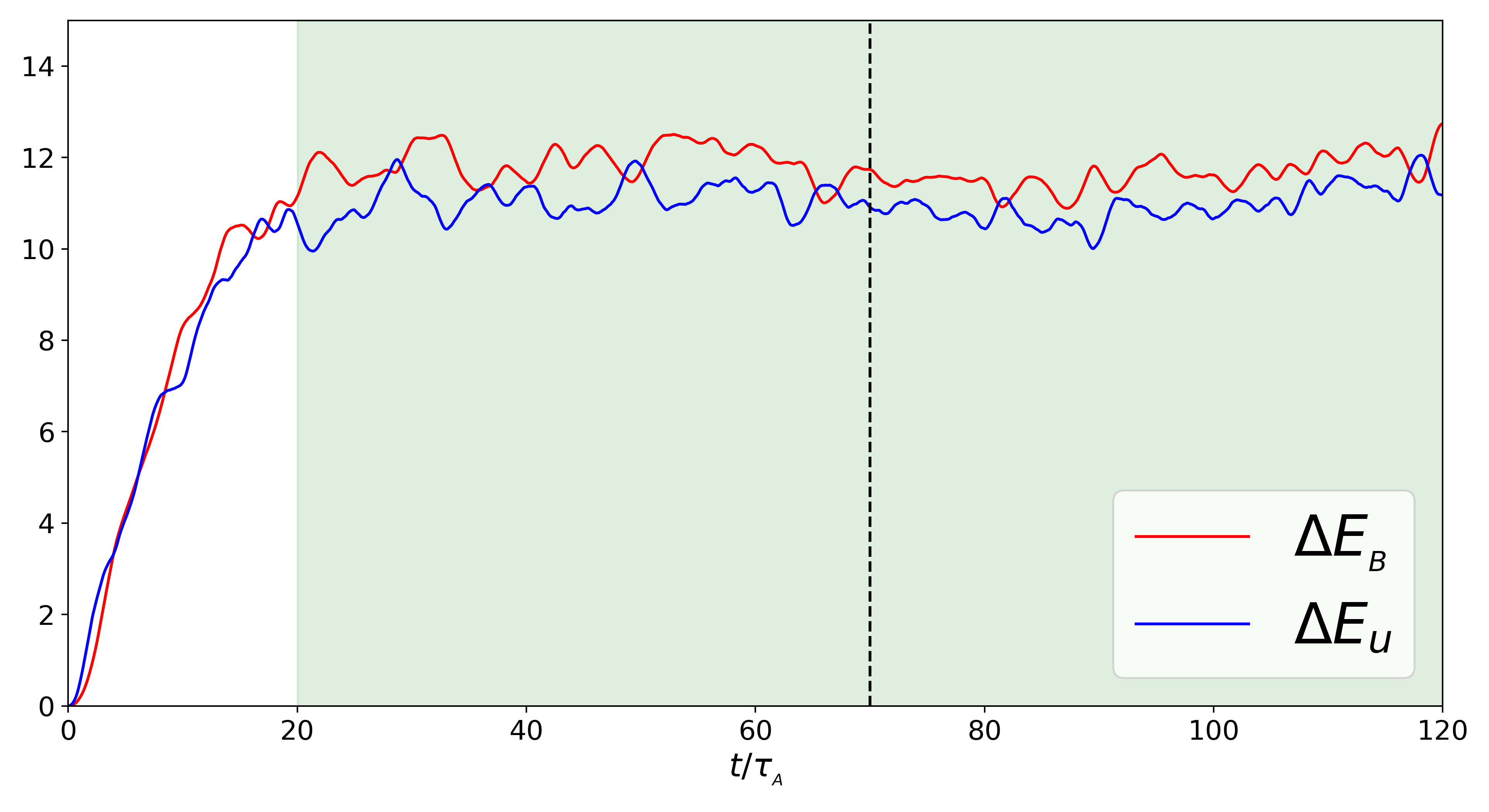}
\caption{Temporal evolution of magnetic and kinetic energy variations $\Delta E_{_B}$ and $\Delta E_u$. The green shaded area indicates the time interval used for the space-time analysis.}
\label{time_hst}
\end{figure}
%%%%%%%%%%%%%%%%%%%%%%%%%%%%%%%%%%%%%%%%%%%%%%%%%%

\begin{figure*}[t]
\centering
\subfloat{
\includegraphics[width=0.32\linewidth]{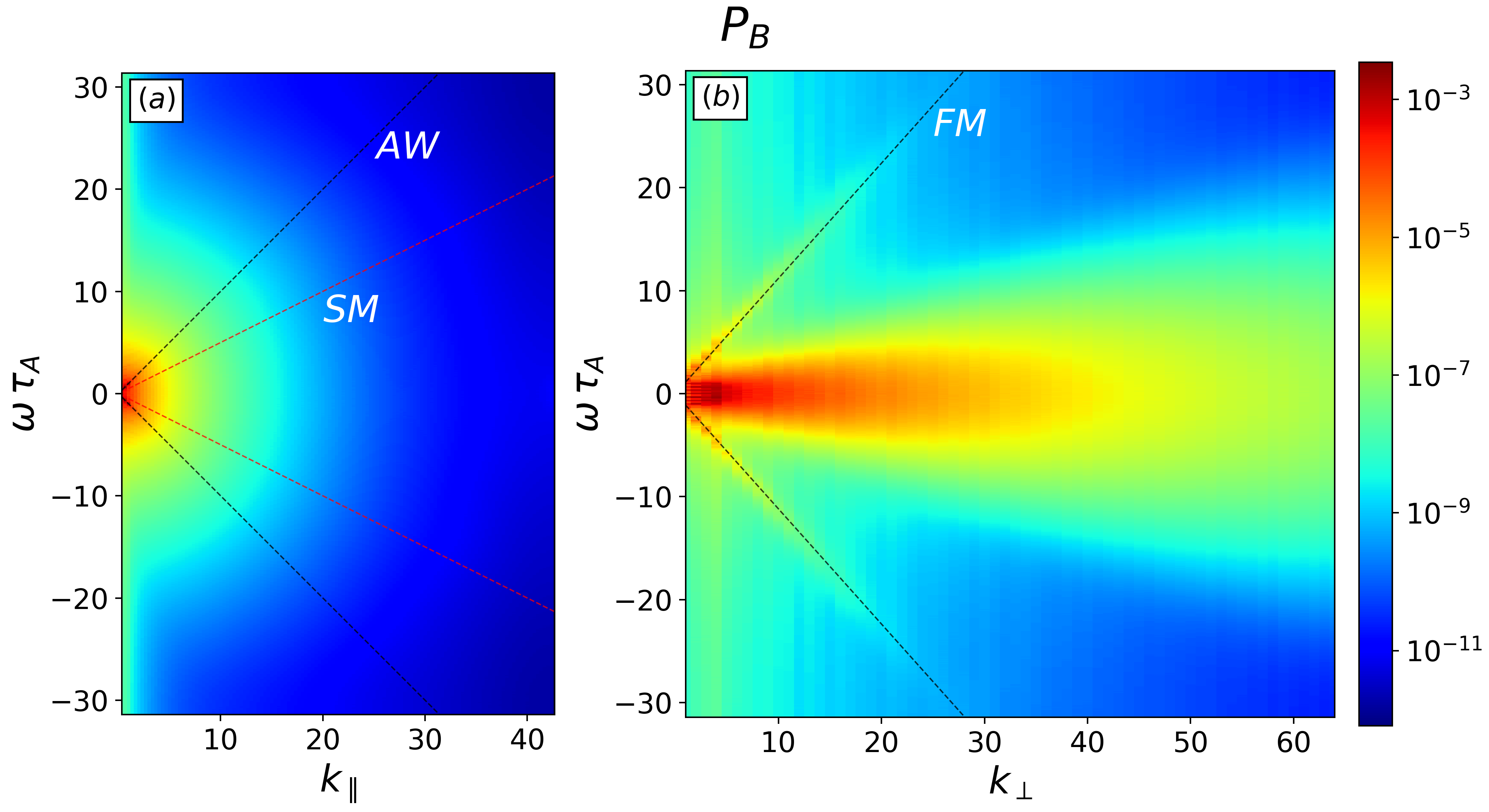}
}
\subfloat{
\includegraphics[width=0.32\linewidth]{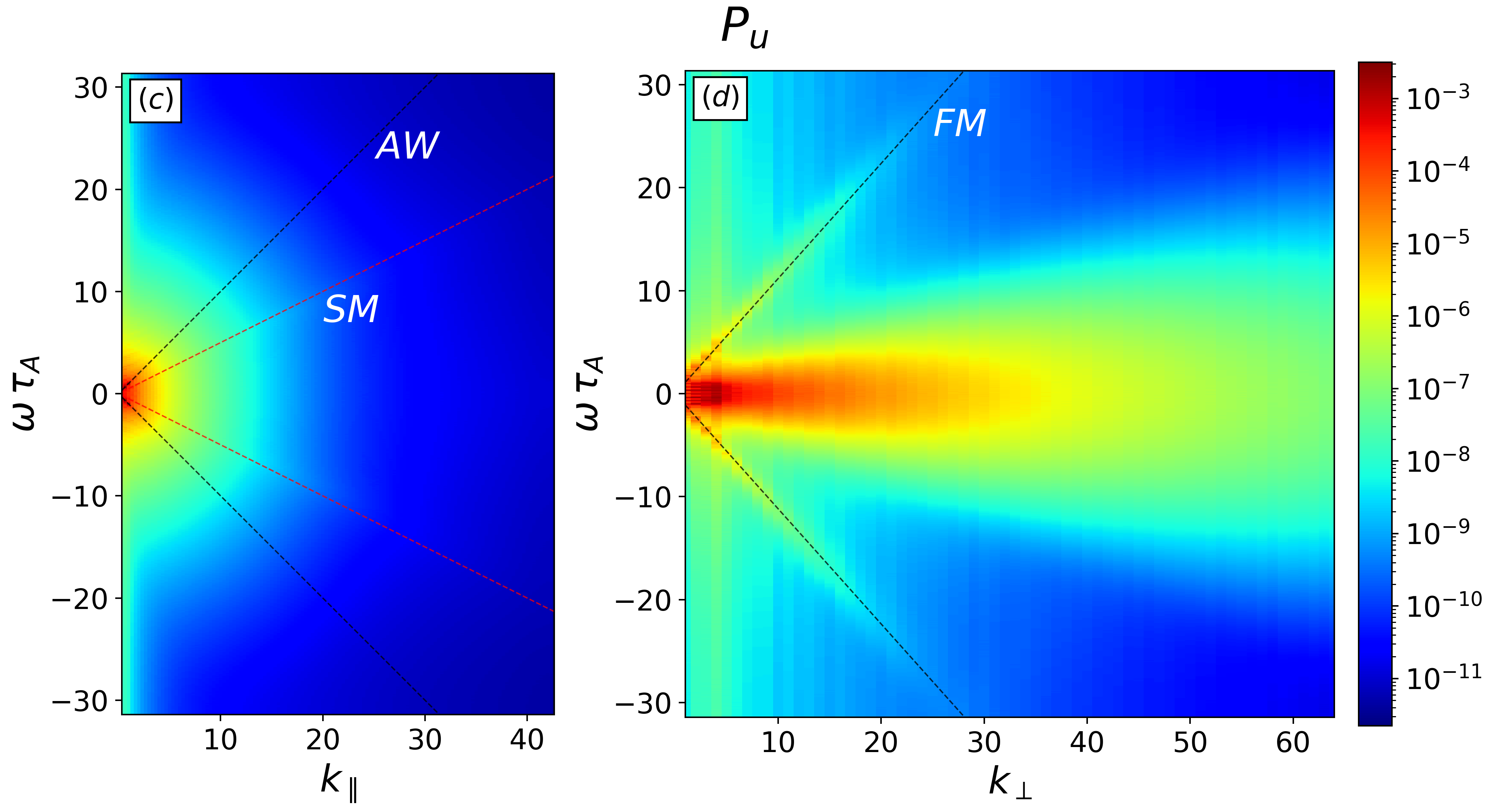}
}
\subfloat{
\includegraphics[width=0.32\linewidth]{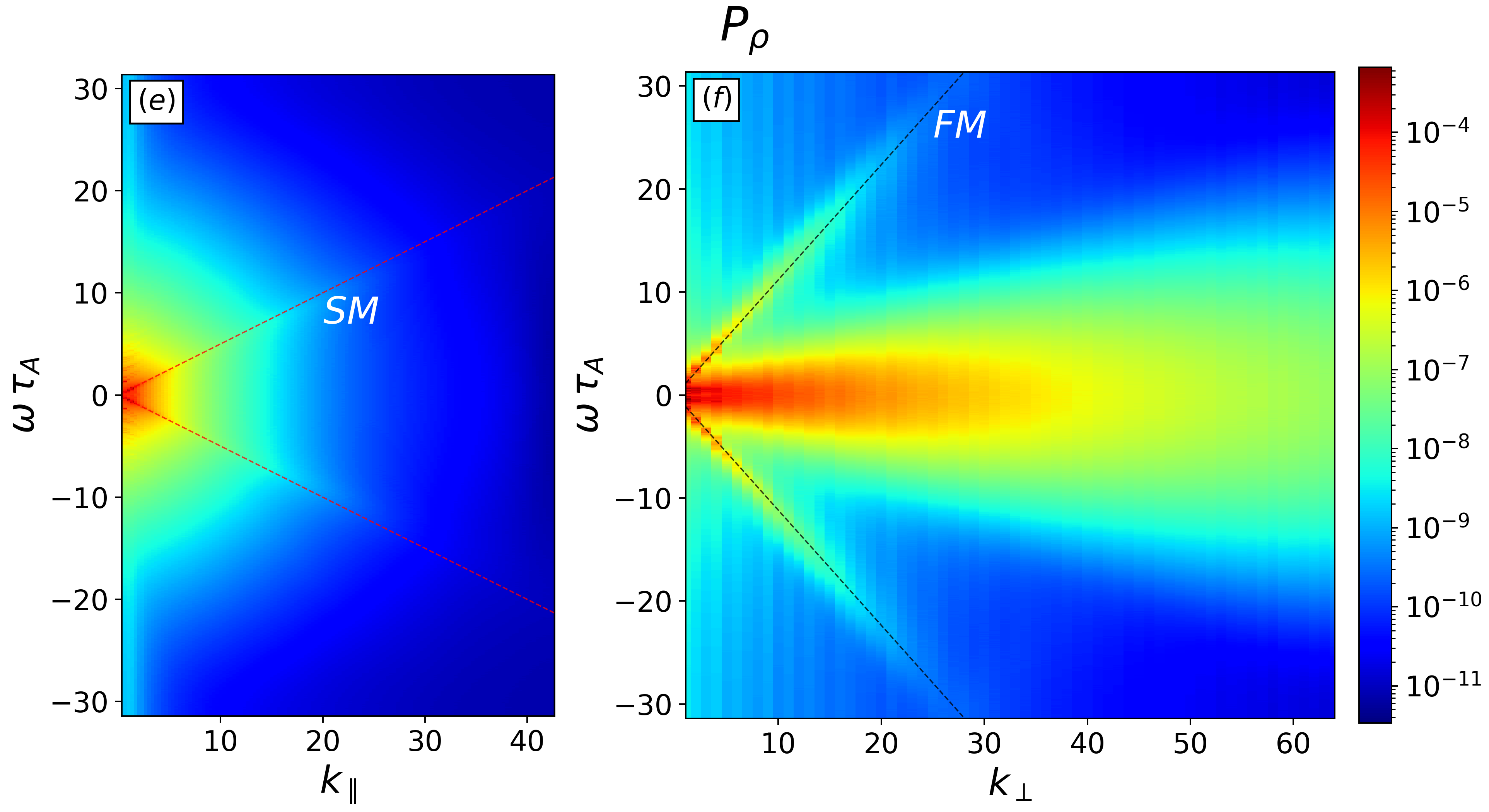}
}
\caption{$(k_{\parallel},\,\omega)$ and $(k_{\perp},\,\omega)$ projections of magnetic field, velocity and density spectra $P_{_B}$ (a)-(b), $P_u$ (c)-(d), and $P_{\rho}$ (e)-(f). Dashed lines indicate the the dispersion relations of Aflvén wave (AW) and slow modes (SM) for parallel propagation ($k_{\perp}\!=\!0$), and of fast modes (FM) for perpendicular propagation ($k_{\parallel}\!=\!0$).}
\label{4DFFT}
\end{figure*}

Figure~\ref{time_hst} shows the temporal evolution of magnetic and kinetic energy variations, $\Delta E_{_B}\!=\!E_{_B}(t)\!-\!E_{_B}(0)$ and $\Delta E_u\!=\!E_u(t)\!-\!E_u(0)$ (in $\rho_0\,c_{_A}^2\,L_0^3$ units). Both energies quickly grow and saturate, after $20\,\tau_{_A}$. We consider interval $T\!=\!\left[20\,\tau_{_A},\,120\,\tau_{_A}\right]$ for our analysis (green shaded area), when turbulence is fully developed and nearly stationary. Figures~\ref{4DFFT}(a)-(b) show $(k_{\parallel},\,\omega)$ and $(k_{\perp},\,\omega)$ projections of the magnetic field space-time Fourier spectrum $P_{_B}$, calculated over interval $T$, with simulation data sampled every $\Delta t\!=\!0.1\,\tau_{_A}$. $(k_{\parallel},\,\omega)$ and $(k_{\perp},\,\omega)$ projections are calculated as $P_{_B}(k_{\parallel},\,\omega) = \int P_{_B}(k_{\perp},\,k_{\parallel},\,\omega) \, dk_{\perp}$ and $P_{_B}(k_{\perp},\,\omega) = \int P_{_B}(k_{\perp},\,k_{\parallel},\,\omega) \, dk_{\parallel}$. The corresponding projections of the velocity spectrum $P_u$ are shown in Fig.~\ref{4DFFT}(c)-(d). We see that energy is distributed into concentric shells in $(k_{\parallel},\,\omega)$ space, quickly falling off toward large $k_{\parallel}$ and $\omega$. $(k_{\perp},\,\omega)$ projections also show that most energy is concentrated around small $\omega$, but with a wider distribution in $k_{\perp}$, up to relatively high wavenumbers. Despite our turbulent driver being incompressible, density fluctuations still develop, reaching an rms amplitude of $\delta \rho_{rms}/\rho_0\!\simeq\!0.08$ at fully developed turbulence. Figures~\ref{4DFFT}(e)-(f) show $(k_{\parallel},\,\omega)$ and $(k_{\perp},\,\omega)$ projections of the density spectrum $P_{\rho}$. Similarly to magnetic and velocity fluctuations, density fluctuations are concentrated around low $\omega$, exhibiting a strong $k_{\perp}\!>\!k_{\parallel}$ anisotropy. Overall, $(k,\,\omega)$ spectra of magnetic, velocity and density fluctuations do not follow the dispersion relations (dashed lines) of Alfvén waves (AW), slow modes (SM), and fast modes (FM), with most energy stored in low $\omega$, low $k_{\parallel}$ fluctuations, consistently with previous numerical works \citep{gan2022existence} and SW observations \citep{zhao2023observations}. 

%%%%%%%%%%%%%%%%%%%%%%%%%%%%%%%%%%%%%%%%%%%%%%%%%%
%%%%%%%%%%%%%%%%%%%%%%%%%%%%%%%%%%%%%%%%%%%%%%%%%%

\begin{figure*}[t]
\centering
\subfloat{
\includegraphics[width=0.32\linewidth]{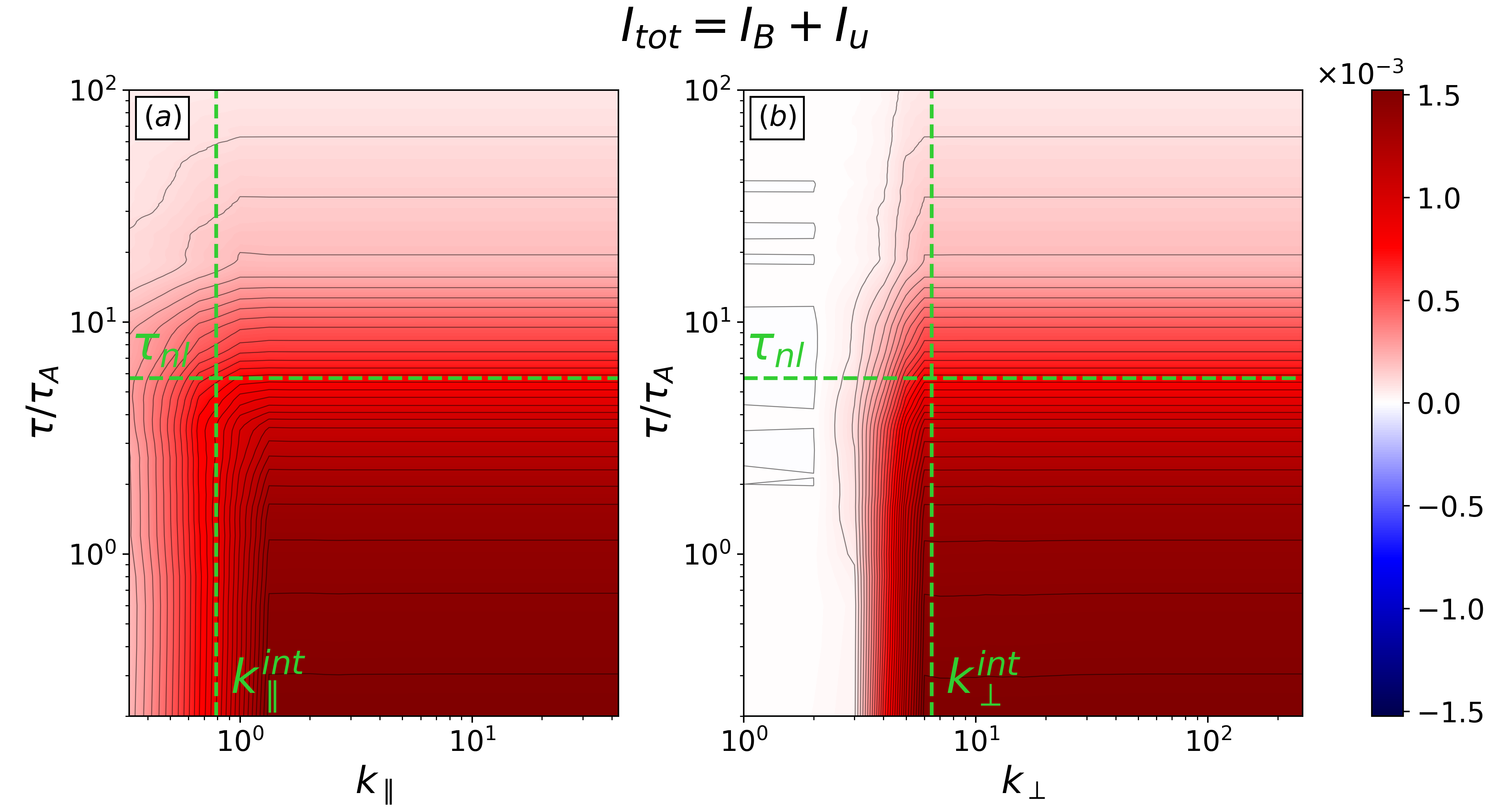}
}
\subfloat{
\includegraphics[width=0.32\linewidth]{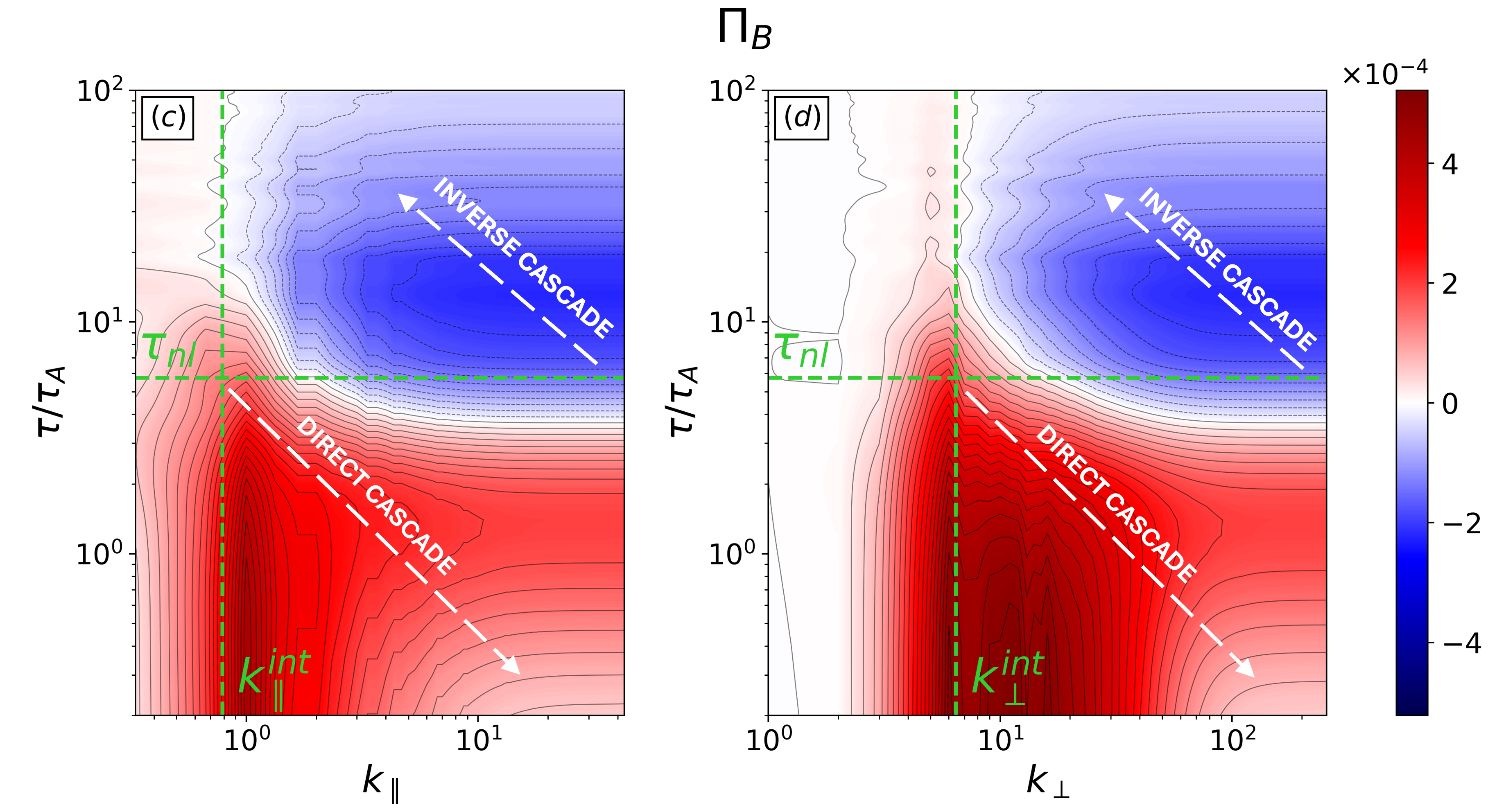}
}
\subfloat{
\includegraphics[width=0.32\linewidth]{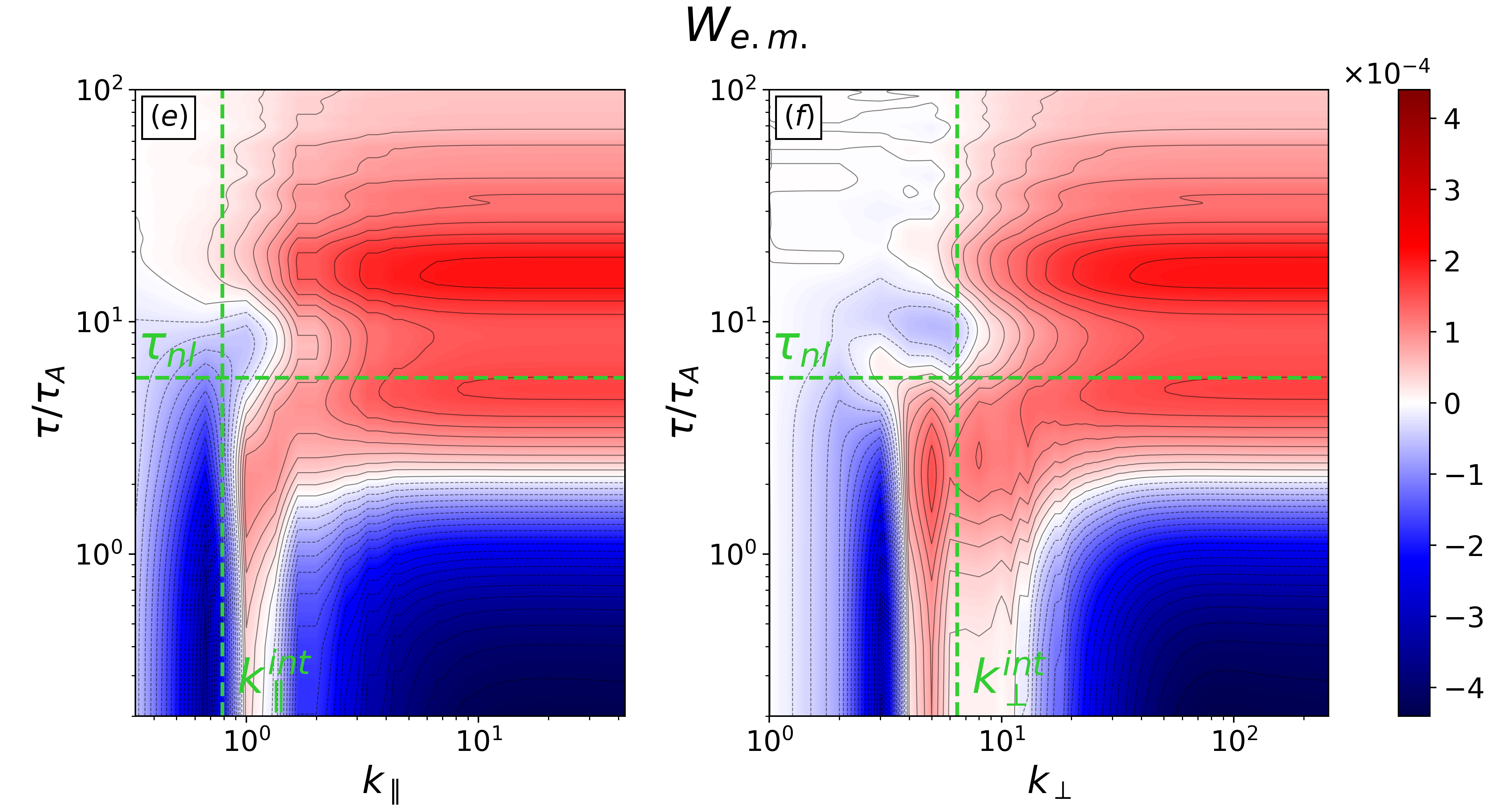}
}
\\
\subfloat{
\includegraphics[width=0.32\linewidth]{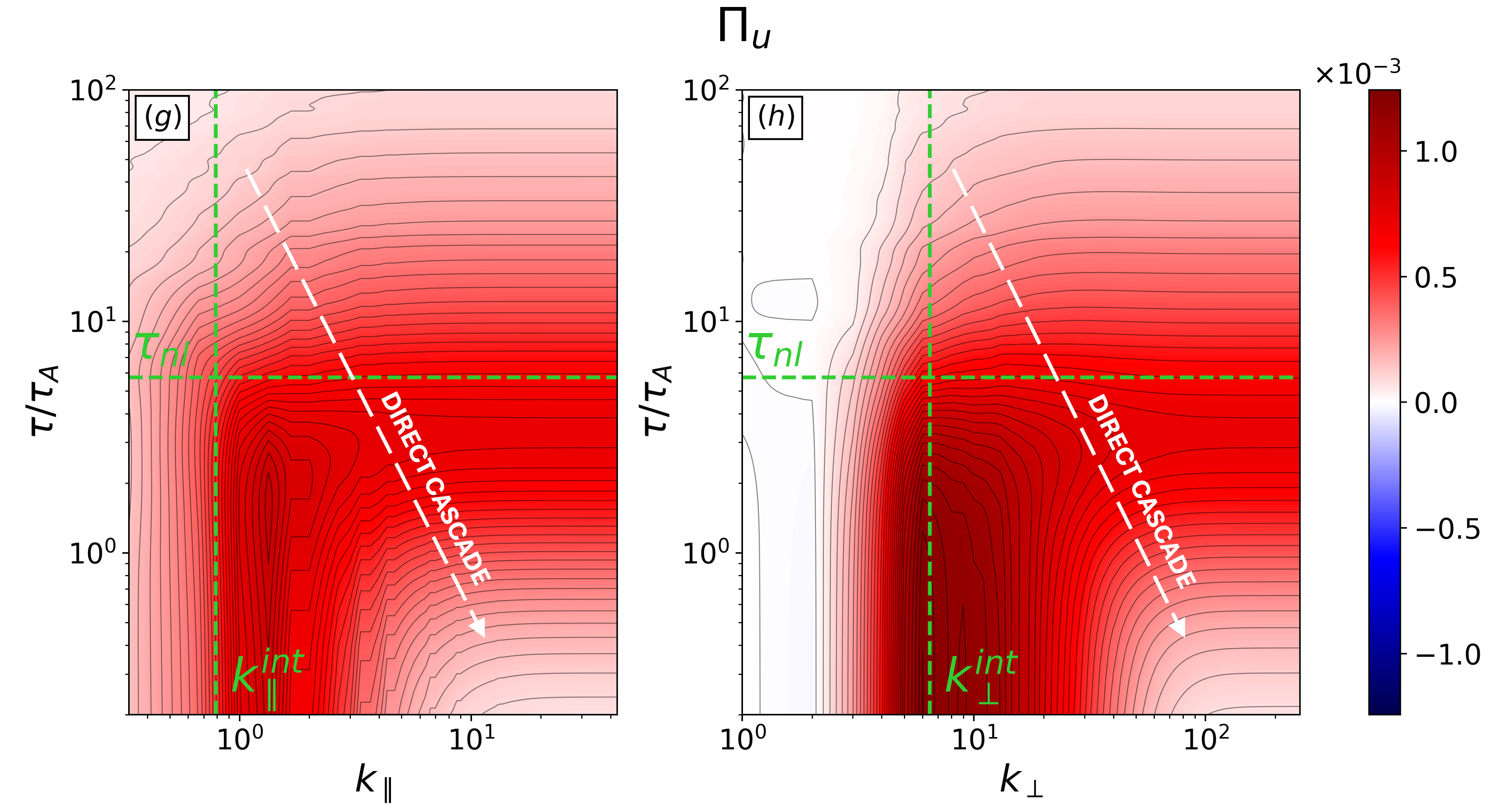}
}
\subfloat{
\includegraphics[width=0.32\linewidth]{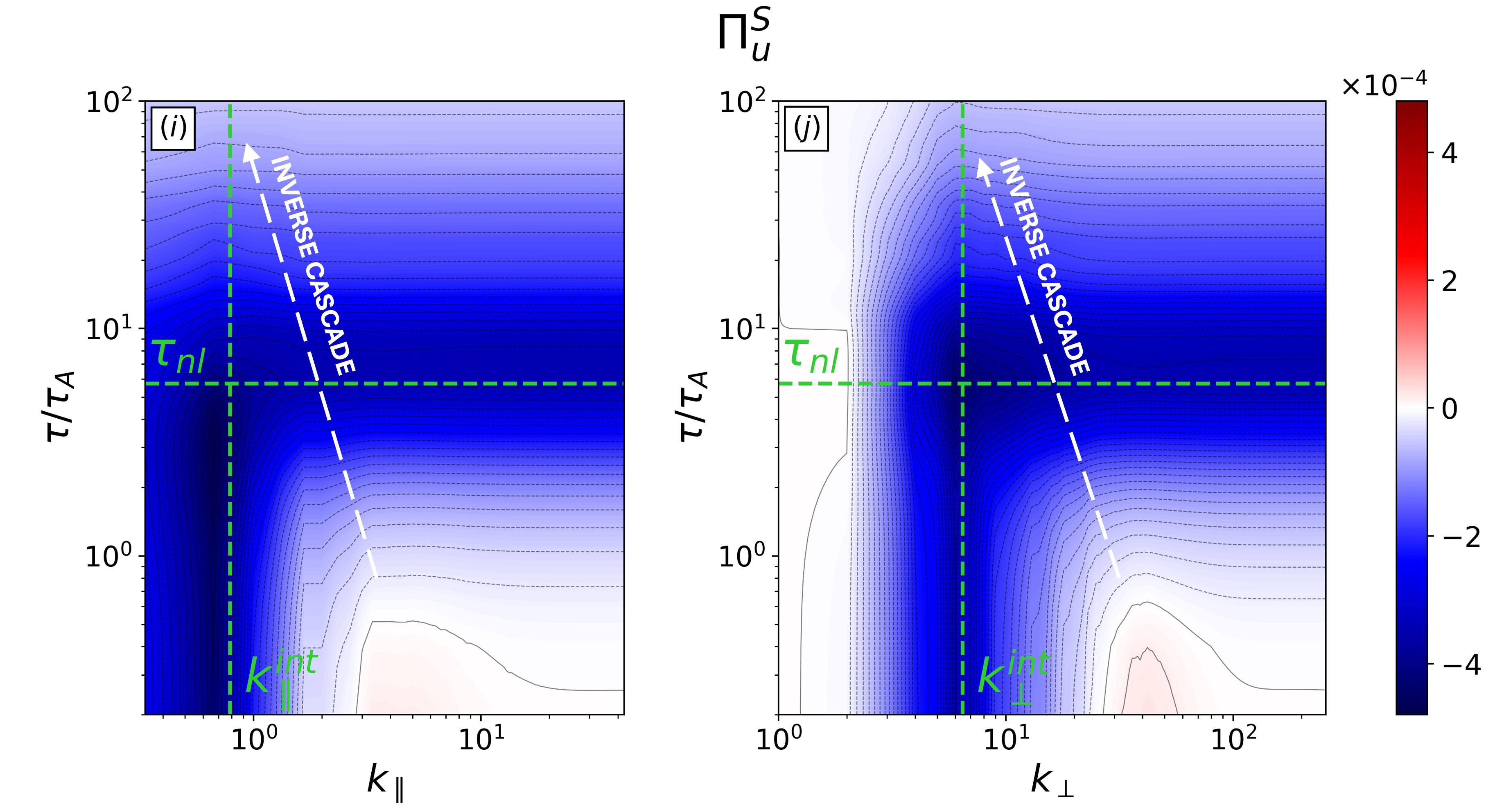}
}
\subfloat{
\includegraphics[width=0.32\linewidth]{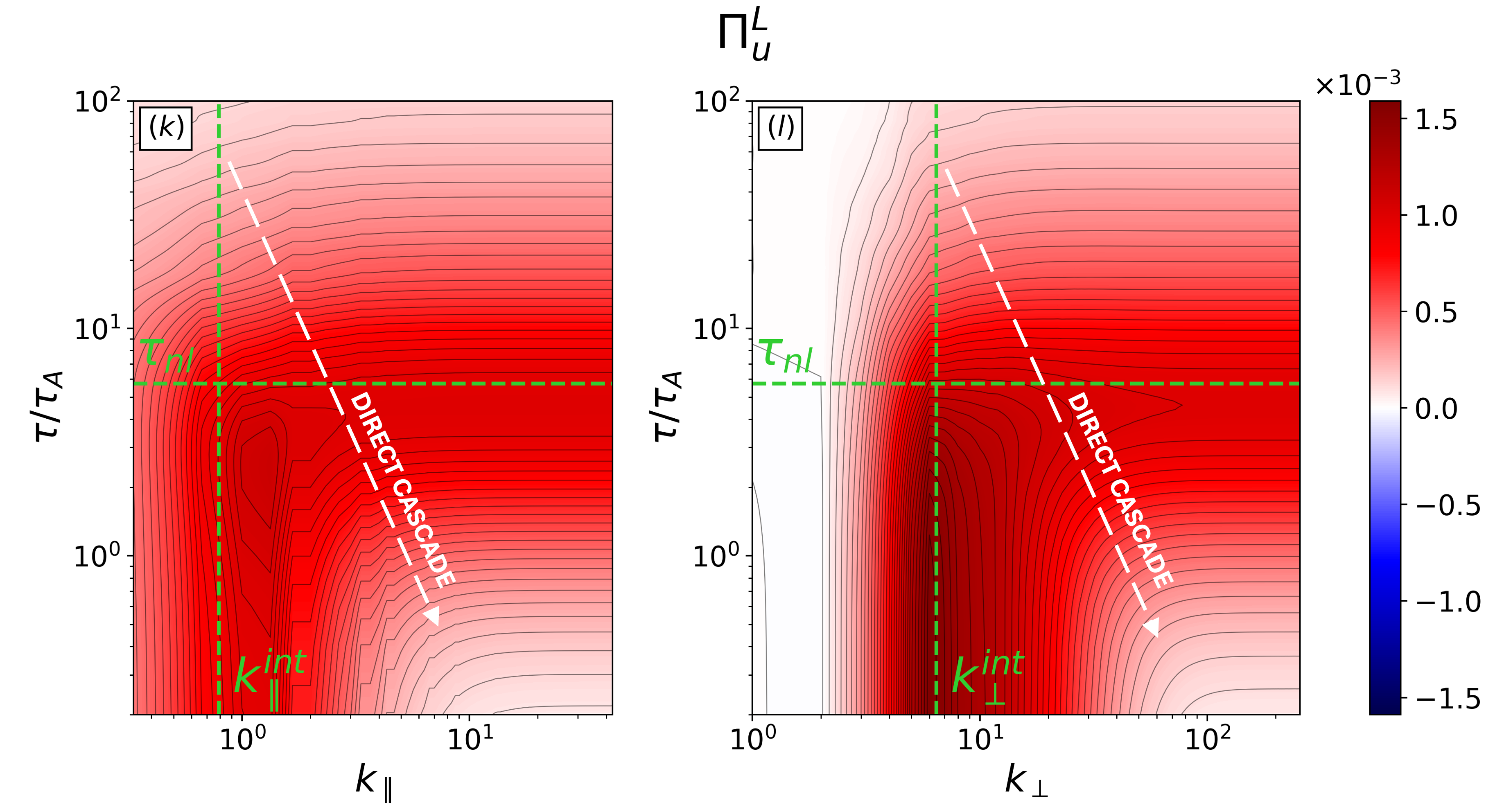}
}
\\
\subfloat{
\includegraphics[width=0.32\linewidth]{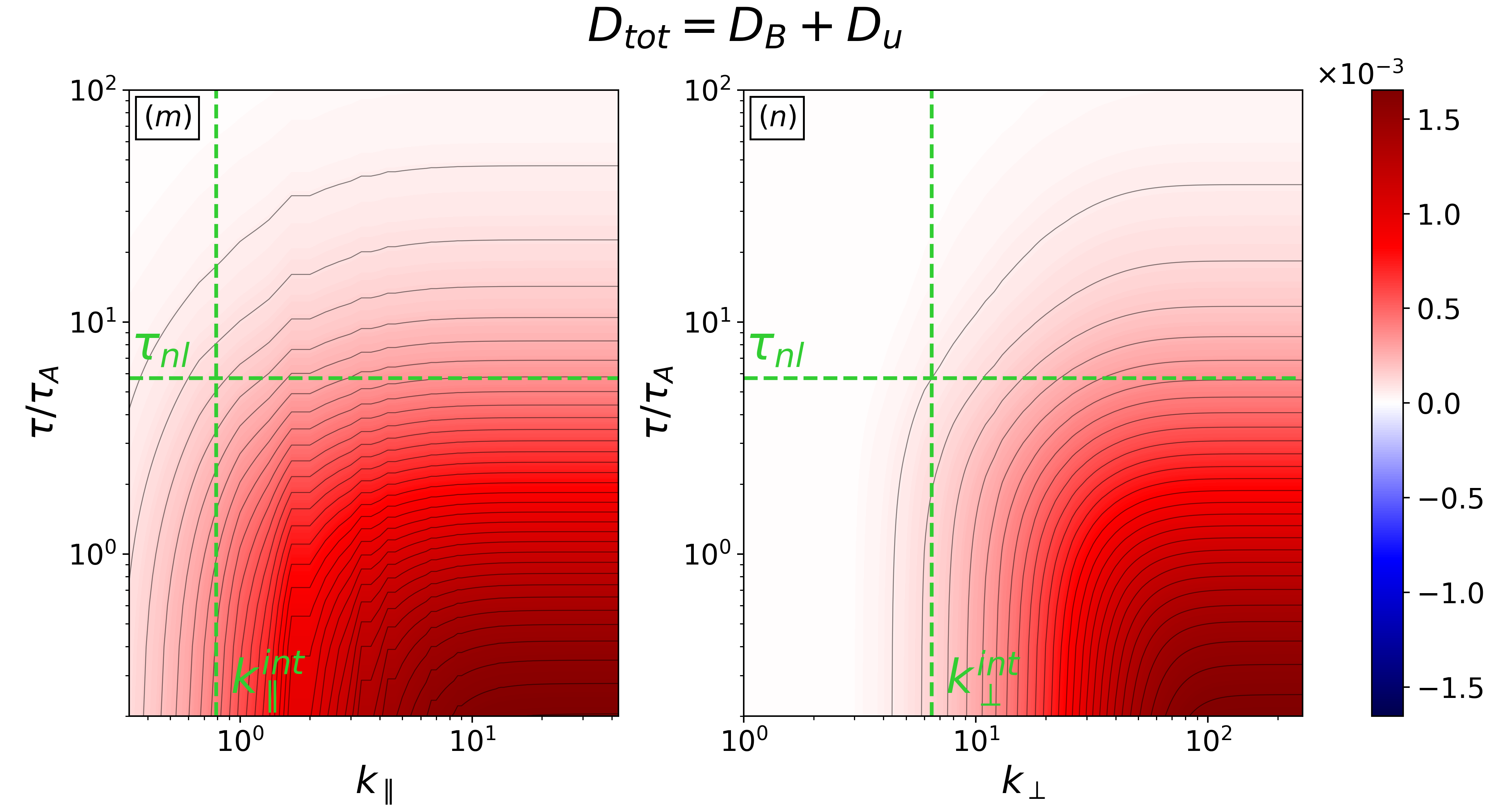}
}
\subfloat{
\includegraphics[width=0.32\linewidth]{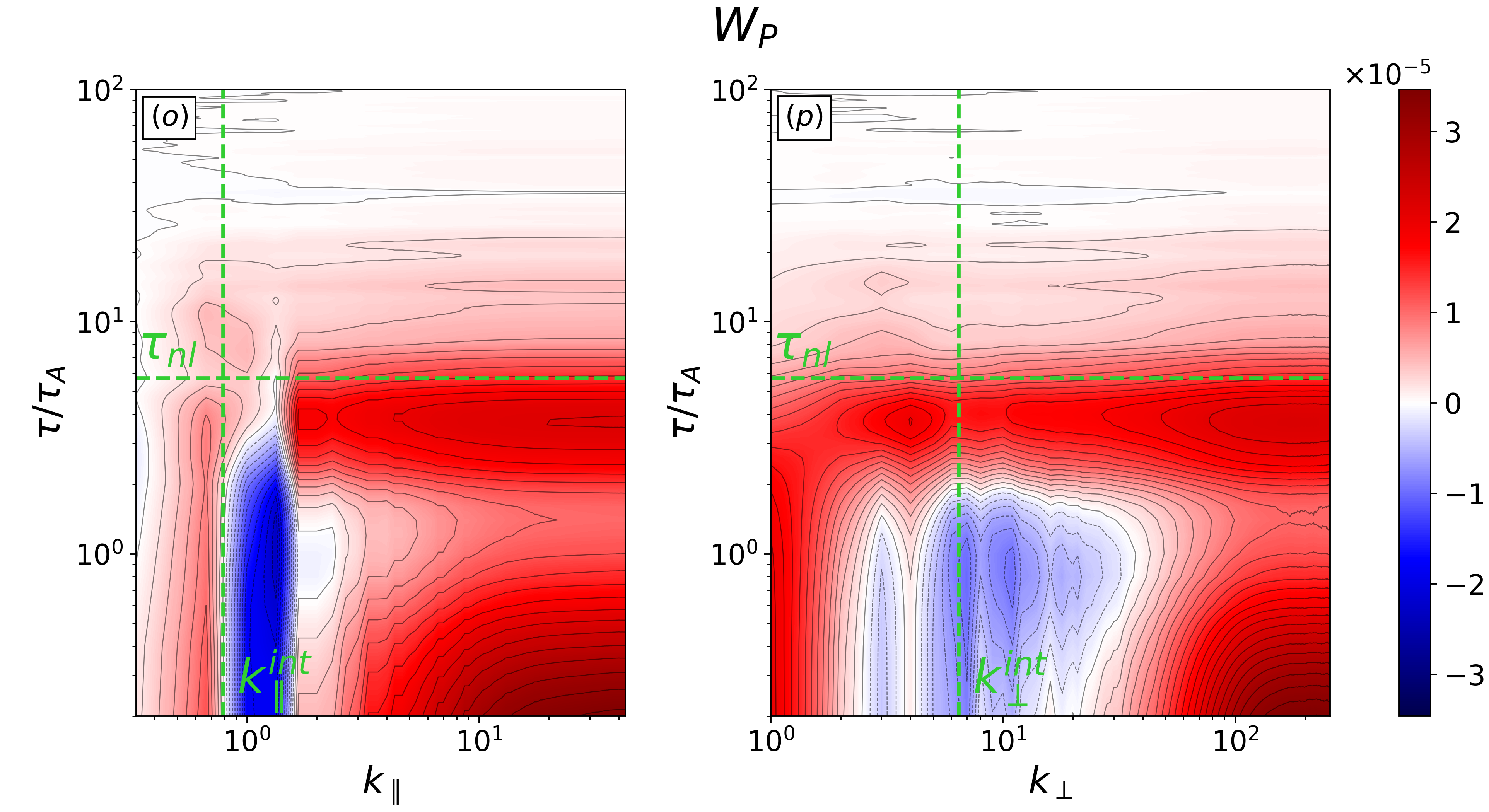}
}
\subfloat{
\includegraphics[width=0.32\linewidth]{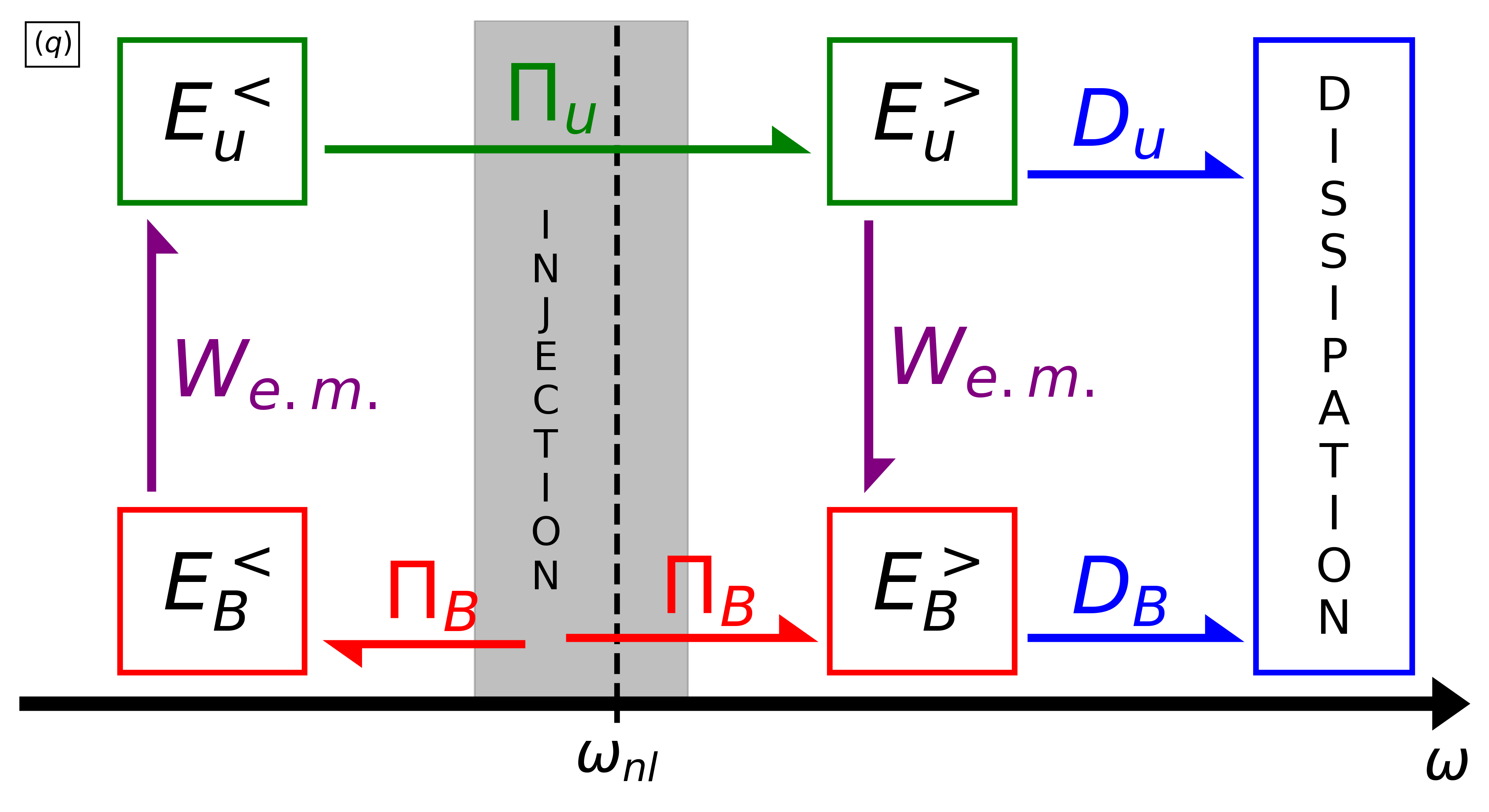}
}
\caption{$(k_{\parallel},\,\tau)$ and $(k_{\perp},\,\tau)$ projections of $I_{_B}+I_u$ (a)-(b), $\Pi_{_B}$ (c)-(d), $W_{e.m.}$ (e)-(f), $\Pi_u$ (g)-(h), $\Pi_u^S$ (i)-(j), $\Pi_u^L$ (k)-(l), $D_{_B}+D_u$ (m)-(n), and $W_{_P}$ (o)-(p), with vertical green dashed lines indicating parallel and perpendicular integral scales $k_{\parallel}^{int}$ and $k_{\perp}^{int}$, while horizontal green dashed lines represent the nonlinear time $\tau_{nl}$. Schematic representation of the energy cascade in frequency space (q).}
\label{4DCG}
\end{figure*}

We investigate the origin of such spectral features using our spatio-temporal CG method, tracking energy transfers from injection to dissipation in $(k_{\perp},\,k_{\parallel},\,\tau)$ space (with $\tau\!\simeq\!2\pi/\omega$). The goal is understanding how low $\omega$ (large $\tau$) fluctuations develop, and their role in the turbulent cascade. We analyze two projections of ETCs in Eqs.~\ref{ETC}: $(k_{\parallel},\,\tau)$ projections are obtained by filtering in $k_{\parallel}$ and $\tau$, retaining all $k_{\perp}$; $(k_{\perp},\,\tau)$ projections are calculated by filtering in $k_{\perp}$ and $\tau$, keeping all $k_{\parallel}$. For the temporal filtering, we center the boxcar kernel $G_{\tau}$ at $t\!=\!70\,\tau_{_A}$ (vertical dashed line in Fig.~\ref{time_hst}), varying its width $\tau$ from $0.2\,\tau_{_A}$ to $100\,\tau_{_A}$, covering the whole interval $T$. ETCs are shown in Fig.~\ref{4DCG}, with black lines indicating their isocontours. As $I_{_B}$ and $I_u$ exhibit analogous $(k_{\perp},\,k_{\parallel},\,\tau)$ dependence and comparable amplitude, we analyze the total energy injection rate $I_{tot}\!=\!I_{_B}\!+\!I_u$. For the same reason, we show the total dissipation rate $D_{tot}\!=\!D_{_B}\!+\!D_u$. $I_{_B}$, $I_u$, $D_{_B}$ and $D_u$ are shown separately in Supplemental Material. For each ECTs, both $(k_{\parallel},\,\tau)$ and $(k_{\perp},\,\tau)$ projections exhibit the same qualitative features, but they are not isotropic in $(k_{\parallel},\,k_{\perp})$ space. The ETCs anisotropy causes the $k_{\perp}\!>\!k_{\parallel}$ anisotropy observed in spectra \citep{du2023anisotropic,hellinger2024anisotropy}. As a reference, we show the parallel and perpendicular integral scales $k_{\parallel}^{int}$ and $k_{\perp}^{int}$ (vertical green dashed lines) \citep{eswaran1988examination}
\begin{equation}
\begin{gathered}
(k_{\perp}^{int},\,k_{\parallel}^{int}) = \frac{ \int\int\int \, (k_{\perp},\,k_{\parallel}) \, P_{u} \, d k_{\perp} d k_{\parallel} d \omega }{ \int\int\int P_{u} \, d k_{\perp} d k_{\parallel} d \omega },
\end{gathered}
\end{equation}
representing the scale of energy containing fluctuations, and the nonlinear time (horizontal green dashed lines) $\tau_{nl} = 2 \pi/(k_{\perp}^{int}\,\delta u_{rms})$. Starting from $I_{tot}$, Fig.~\ref{4DCG}(a)-(b), we see it is positive and increases with wavenumber, up to $k_{\parallel}\!\simeq\!1.3$ and $k_{\perp}\!\simeq\!6$, after which it saturates, indicating no contribution from larger wavenumbers. Regarding its $\tau$ dependence, we note that $I_{tot}$ is not peaked at the driving time $2\pi/\omega_0\!\simeq\!7.8\,\tau_{_A}$, but exhibits a broad frequency response, growing from $\tau\!\simeq\!15$ to $\tau\!\simeq\!1.5$, while being almost constant elsewhere. This broadening is determined by the decorrelation rate $\gamma_0$ of the LA driver, typically chosen to be slightly smaller than $\omega_0$. Different drivers may alter the $(k,\,\omega)$ distribution of injected perturbations, but we choose the LA approach since it has been proven to successfully reproduce key features of SW turbulence \citep{tenbarge2012evidence,tenbarge2014oscillating}. 

%%%%%%%%%%%%%%%%%%%%%%%%%%%%%%%%%%%%%%%%%%%%%%%%%%
%%%%%%%%%%%%%%%%%%%%%%%%%%%%%%%%%%%%%%%%%%%%%%%%%%

Energy injected by $I_{tot}$ is transferred to other scales by the cascade terms $\Pi_{_B}$ and $\Pi_{u}$. Starting from $\Pi_{_B}$, Fig.~\ref{4DCG}(c)-(d), we find it is negative for $\tau\!\gtrsim\!\tau_{nl}$ and wavenumbers larger than $k_{\parallel}^{int}$ and $k_{\perp}^{int}$, while being positive for $\tau\!<\!\tau_{nl}$, where it peaks around integral scales. This indicates a frequency space bifurcation in the magnetic energy cascade, with large $\tau$ (low $\omega$) fluctuations exhibiting an inverse cascade\footnote{By \enquote{inverse cascade} we mean energy transfer from high to low frequencies. Other aspects typically associated with the traditional concept of \enquote{cascade}, such as its locality, are not addressed here and will be investigated in future studies.} toward even lower frequencies and smaller wavenumbers in $(k,\,\tau)$ space, while small $\tau$ (high $\omega$) magnetic modes undergo a direct cascade, with the cascade rate being stronger around $k_{\parallel}^{int}$ and $k_{\perp}^{int}$, in the inertial range. Hence, part of the injected magnetic energy is transferred to low $(k,\,\omega)$ fluctuations, while another fraction is transferred to high $(k,\,\omega)$ modes. We first follow the path taken by low $\omega$ magnetic fluctuations. Magnetic energy piling up at large $\tau$ because of the inverse cascade can be converted into kinetic energy by $W_{e.m.}$. Figures~\ref{4DCG}(e)-(f) show that $W_{e.m.}$ is positive in the same range where $\Pi_{_B}$ is negative, for $\tau\!\gtrsim\!\tau_{nl}$ and wavenumbers larger than $k_{\parallel}^{int}$ and $k_{\perp}^{int}$, and their absolute magnitudes are similar, indicating that low $\omega$ magnetic energy is converted into low $\omega$ kinetic energy. Such balance between the inverse magnetic energy cascade and a low $\omega$ energy sink, $W_{e.m.}$ in this case, is expected and has to occur in order for turbulence to reach a quasi-stationary state. The combination of the inverse magnetic energy cascade and the magnetic-to-kinetic energy conversion at low $\omega$, implies an accumulation of magnetic and kinetic energy at small frequencies, explaining the origin of low $\omega$ fluctuations in energy spectra. Unlike $\Pi_{_B}$, the kinetic energy cascade rate $\Pi_{u}$, Fig.~\ref{4DCG}(g)-(h), is positive at all $(k,\,\tau)$ scales, peaking around integral scales and at $\tau$ slightly smaller than $\tau_{nl}$, indicating a direct kinetic energy cascade, regardless of the frequency. Combining $\Pi_{_B}$, $W_{e.m.}$ and $\Pi_u$, we thus find that low $\omega$ magnetic fluctuations contribute to driving turbulence, providing magnetic energy that is first converted into low $\omega$ kinetic energy by $W_{e.m.}$, and finally transferred to higher $(k,\,\omega)$ fluctuations by $\Pi_{u}$. Since $W_{e.m.}$ is negative for large $k$ and small $\tau$, Fig.~\ref{4DCG}(e)-(f), part of the small scale kinetic energy resulting from the direct cascade is converted to small scale magnetic energy. 

Additional insights on the turbulent cascade are obtained decomposing $\Pi_u$ into $\Pi_u^S$ and $\Pi_u^L$, in Fig.~\ref{4DCG}(i)-(l). We find that $\Pi_u^S$ is negative over most of $(k_{\perp},\,k_{\parallel},\,\tau)$ space (except at small scales, where it is weakly positive), indicating an inverse cascade. This is strikingly different from hydrodynamic turbulence, where $\Pi_u^S$ is typically positive, being the only cascade term \citep{frisch1995turbulence}. Conversely, $\Pi_u^L$ is positive at all scales, and stronger in amplitude than $\Pi_u^S$, thus producing a net direct kinetic energy cascade when combined with $\Pi_u^S$. Hence, the direct kinetic energy cascade in MHD turbulence is mainly driven by e.m. interactions between the small scale Lorentz force and large scale velocities, quantified by $\Pi_u^L$, while the hydrodynamic cascade term $\Pi_u^S$ opposes the direct $\Pi_u^L$ cascade. 

Magnetic and kinetic energy cascading to small scales is eventually dissipated. Figures~\ref{4DCG}(m)-(n) show the dissipation rate $D_{tot}$ to be weak at large scales, growing to large positive values toward high $k$ and small $\tau$, indicating that dissipation mainly affects high frequencies and wavenumbers. Consequently, low $\omega$ fluctuations resulting from the inverse cascade are not dissipated. Turbulent SW dissipation is mainly caused by kinetic processes \citep{alexandrova2013solar}, rather than by collisional resistive and viscous effects as in our simulation. However, kinetic scale dissipation is typically mediated by wave-particle interactions \citep{squire2022high} and fast intermittent events, like magnetic reconnection \citep{papini2019can,arro2020statistical}, whose time scales are comparable to particle gyroperiods. Hence, we expect kinetic effects not to affect the dynamics of low $\omega$ fluctuations, whose time scales are much larger than particle gyroperiods \citep{zhao2023observations}. 

Another channel for kinetic energy dissipation is $W_{_P}$, quantifying compressible effects. Figures~\ref{4DCG}(o)-(p) show $W_{_P}$ to be negligibly small for $\tau\!>\!\tau_{nl}$, meaning that low $\omega$ fluctuations are essentially incompressible, while for $\tau\!\lesssim\!\tau_{nl}$ it is negative around integral scales (dilatation), and positive (compression) elsewhere. However, $W_{_P}$ is about one order of magnitude smaller than other ETCs, providing negligible contributions to the global energy balance, consistently with previous studies \citep{yang2016energy}. 

The interplay among all ECTs in frequency space is schematically summarized in Fig.~\ref{4DCG}(q), highlighting the origin and role of low $\omega$ turbulent fluctuations. Energy injection is represented by the gray shaded area around the nonlinear frequency $\omega_{nl}\!\simeq\!2\pi/\tau_{nl}$. Part of the injected magnetic energy exhibits an inverse cascade, driven by $\Pi_{_B}$, causing a pileup of low $\omega$ magnetic energy $E_{_B}^{<}$, and thus of low $\omega$ magnetic fluctuations. Low $\omega$ magnetic energy is transferred to low $\omega$ kinetic energy $E_{u}^{<}$ by the e.m. work $W_{e.m.}$, producing low $\omega$ velocity fluctuations. Low $\omega$ kinetic energy undergoes a direct cascade, driven by $\Pi_u$, toward high $\omega$ kinetic energy $E_{u}^{>}$. Energy flows into high $\omega$ magnetic energy $E_{_B}^{>}$ through two paths: a direct cascade from injection to $E_{_B}^{>}$, driven by $\Pi_{_B}$, and the kinetic-to-magnetic energy transfer driven by $W_{e.m.}$ at high $\omega$, converting $E_{u}^{>}$ into $E_{_B}^{>}$. Finally, $E_{_B}^{>}$ and $E_{u}^{>}$ are dissipated by $D_{_B}$ and $D_u$, respectively. $E_{_B}^{<}$ and $E_{u}^{<}$ are not affected by resistive and viscous dissipation, making low $\omega$ magnetic field and velocity fluctuations a stable reservoir of energy that contributes to driving the turbulent cascade. The fact that dissipation is efficient at high $\omega$, while not affecting low $\omega$, explains the energy distribution observed in magnetic field and velocity spectra, where most energy lies in low $\omega$ modes, while high $\omega$ fluctuations are dissipated, exhibiting a negligible energy content.

The main new mechanism unveiled by our spatio-temporal CG analysis is the frequency space bifurcation in the magnetic energy cascade, with magnetic fluctuations exhibiting an inverse cascade at frequencies $\omega\!<\!\omega_{nl}$, and a direct cascade for $\omega\!>\!\omega_{nl}$. Our results potentially explain the origin of low frequency turbulent fluctuations observed in the SW, showing that low $\omega$ modes are not just a passive product of turbulence, but they also support the energy cascade. Additional elements may contribute to low frequency SW fluctuations, as the presence of long-lived structures originating in the solar chromosphere or corona, later advected into the SW. Nevertheless, our analysis shows that turbulence alone is capable of producing low frequency fluctuations locally, suggesting they may develop in-situ in the SW. Here by \enquote{low frequency fluctuations} we mean the low $\omega$ range in $(k,\,\omega)$ spectra in Fig.~\ref{4DFFT}, also measured in the SW \citep{narita2010magnetic,zhao2023observations}. However, low frequency SW fluctuations cover a wide frequency range, often exhibiting $1/f$ spectra extending about two decades below the turbulence correlation frequency. Such a $\omega$ range is only partially accessible in our simulation, as the largest time scales we resolve are about one order of magnitude larger than $\tau_{nl}$.

Our findings have potential implications for the development of new models of plasma turbulence, including frequency information. The inverse magnetic energy cascade we observe is fundamentally different from the turbulent dynamo. In dynamo configurations, small scale low amplitude magnetic perturbations are amplified and grow in size due to field line stretching induced by large scale hydrodynamic flows, with a wide scale separation between velocity and magnetic fluctuations \citep{schekochihin2004simulations}. This is not the case in our simulation, since both magnetic and velocity perturbations are injected in the same range of scales, with similar amplitudes, mimicking SW turbulence driven by large scale Alfvénic fluctuations. In this scenario, we find that magnetic fluctuations with different $\omega$ contribute differently to turbulence, with a fraction of them exhibiting an inverse cascade, while others undergo a direct cascade.

The turbulent cascade may be influenced by several parameters, including $\beta$ and $\sigma_{_C}$ \citep{biskamp1997nonlinear,lugones2019spatio,schekochihin2022mhd}. To test the generality of our findings, we applied our CG analysis to other two MHD simulations of turbulence, a $(\beta\!=\!18,\,\sigma_{_C}\!\simeq\!0)$ run, representing SW turbulence in the outer heliosphere \citep{fraternale2022exploring}, and a $(\beta\!=\!0.5,\,\sigma_{_C}\!\simeq\!0.73)$ run, reproducing near-Sun SW turbulence, where $\sigma_{_C}$ reaches large values (imbalanced turbulence), as measured by Parker Solar Probe \citep{chen2020evolution}. In both runs, the energy cascade exhibits the same spatio-temporal behavior observed in our main $(\beta\!=\!0.5,\,\sigma_{_C}\!\simeq\!0)$ run. Specifically, an inverse magnetic energy cascade occurs at $\tau\!\gtrsim\!\tau_{nl}$, with magnetic energy being converted into kinetic energy by $W_{e.m.}$ at low $\omega$, complemented by a direct kinetic energy cascade from low to high frequencies (see Supplemental Material). Another parameter that may influence the cascade dynamics is the magnetic helicity $\sigma_{_B}$, a conserved MHD quantity undergoing an inverse cascade in wavenumber space \citep{frisch1975possibility,alexakis2006inverse}. The random phase nature of our magnetic driver $\textbf{F}_{_B}$ implies the injection of magnetic fluctuations with $\sigma_{_B}\!\simeq\!0$, so we argue that the inverse magnetic energy cascade we observe is not related to the inverse $\sigma_{_B}$ cascade. 

We have shown that low $\omega$ magnetic and velocity modes are also associated with density fluctuations, while being nearly incompressible. The question remains to understand the identity of such low $\omega$ incompressible structures. A possible interpretation is that low $\omega$ fluctuations correspond to long-lived flux ropes, typically observed in the SW, where magnetic pressure is balanced by density (and thus pressure) variations \citep{hu2001reconstruction,lui2011grad,pecora2019single}. Investigating the properties of low $\omega$ turbulent fluctuations in terms of their real space structure will be the subject of future studies. 

\begin{acknowledgments}

\vspace{9pt}

G. A. acknowledges useful discussions with Prof. Fabio Bacchini, regarding the implementation of the turbulent forcing and other MHD numerical methods. H.L. acknowledges the support by LANL/LDRD program and DOE FES program.

This research is supported by the NASA grant No. 80NSSC23K0101. Opinions, findings, and conclusions expressed in this work are those of the authors and do not necessarily reflect the views of NASA.

Numerical simulations and data analysis have been performed on Perlmutter at NERSC (USA), under the FES-ERCAP-m4239 project. This research used resources provided by the Los Alamos National Laboratory Institutional Computing Program, which is supported by the U.S. Department of Energy National Nuclear Security Administration under Contract No. 89233218CNA000001.

\end{acknowledgments}

\appendix

\section{\large Supplemental Material}

\section{Coarse grained magnetohydrodynamic equations}

%%%%%%%%%%%%%%%%%%%%%%%%%%%%%%%%%%%%%%%%%%%%%%%%%%
\begin{figure*}[t]
\centering
\subfloat{
\includegraphics[width=0.47\linewidth]{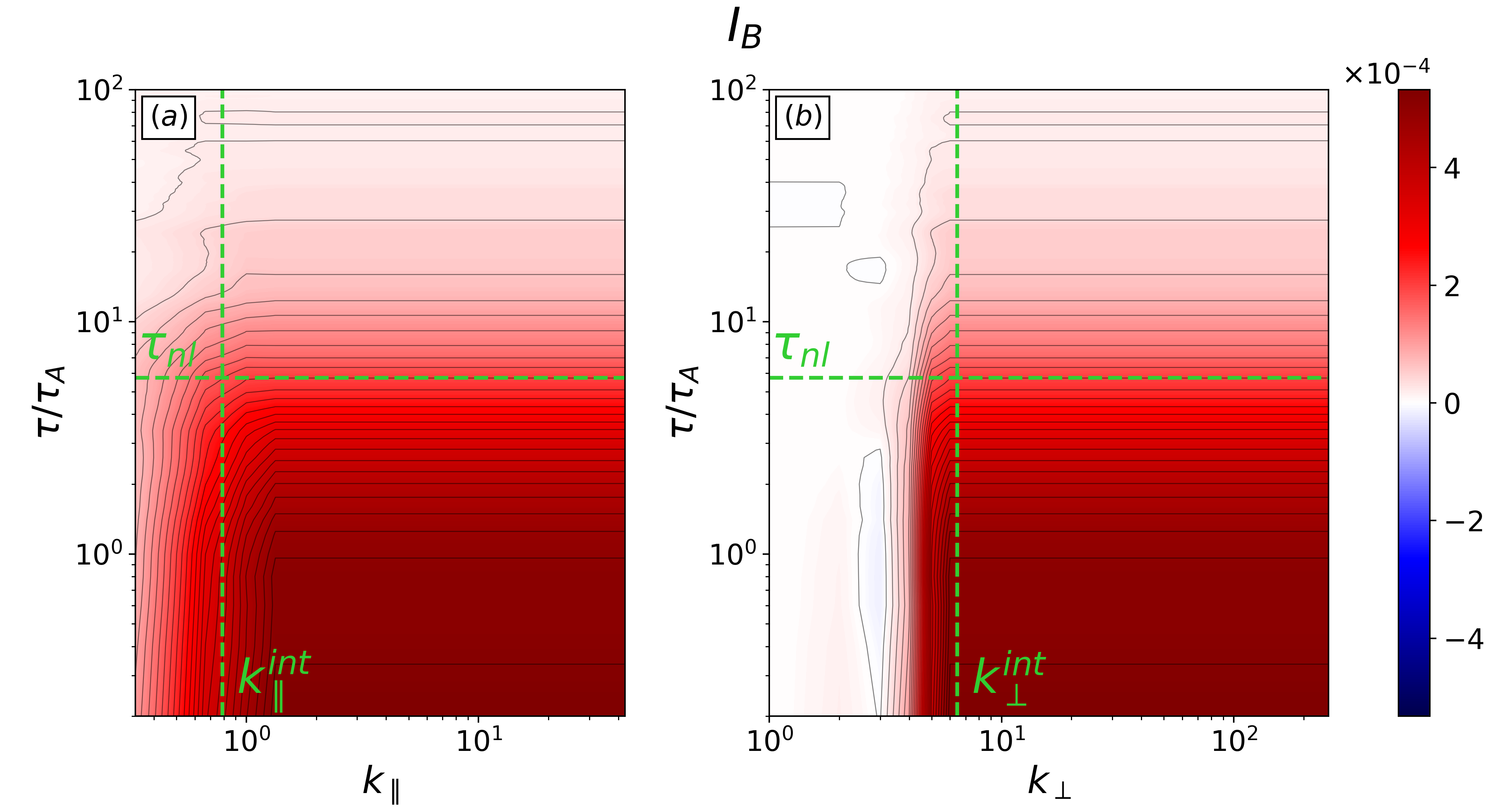}
}
\subfloat{
\includegraphics[width=0.47\linewidth]{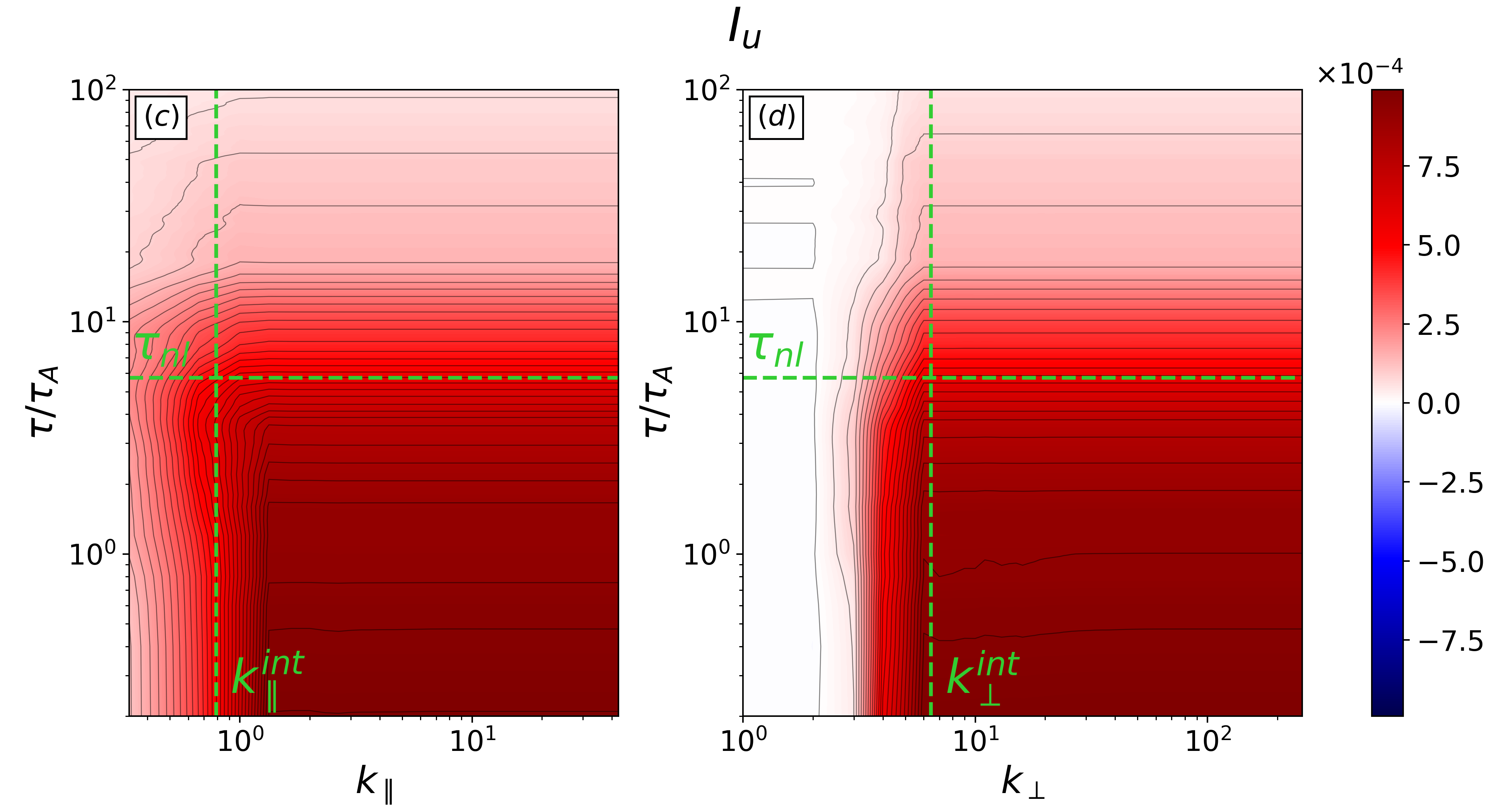}
}
\\
\subfloat{
\includegraphics[width=0.47\linewidth]{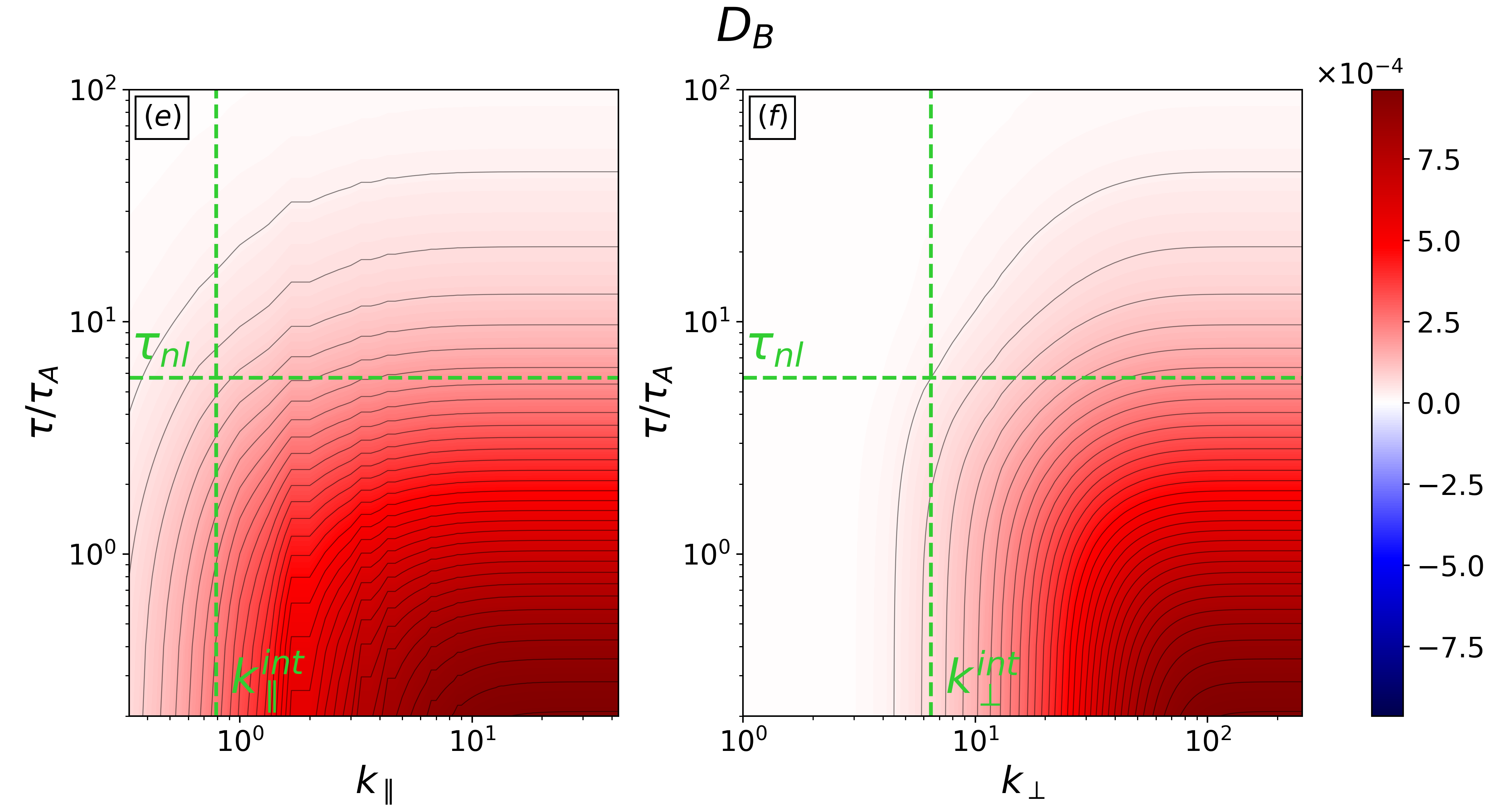}
}
\subfloat{
\includegraphics[width=0.47\linewidth]{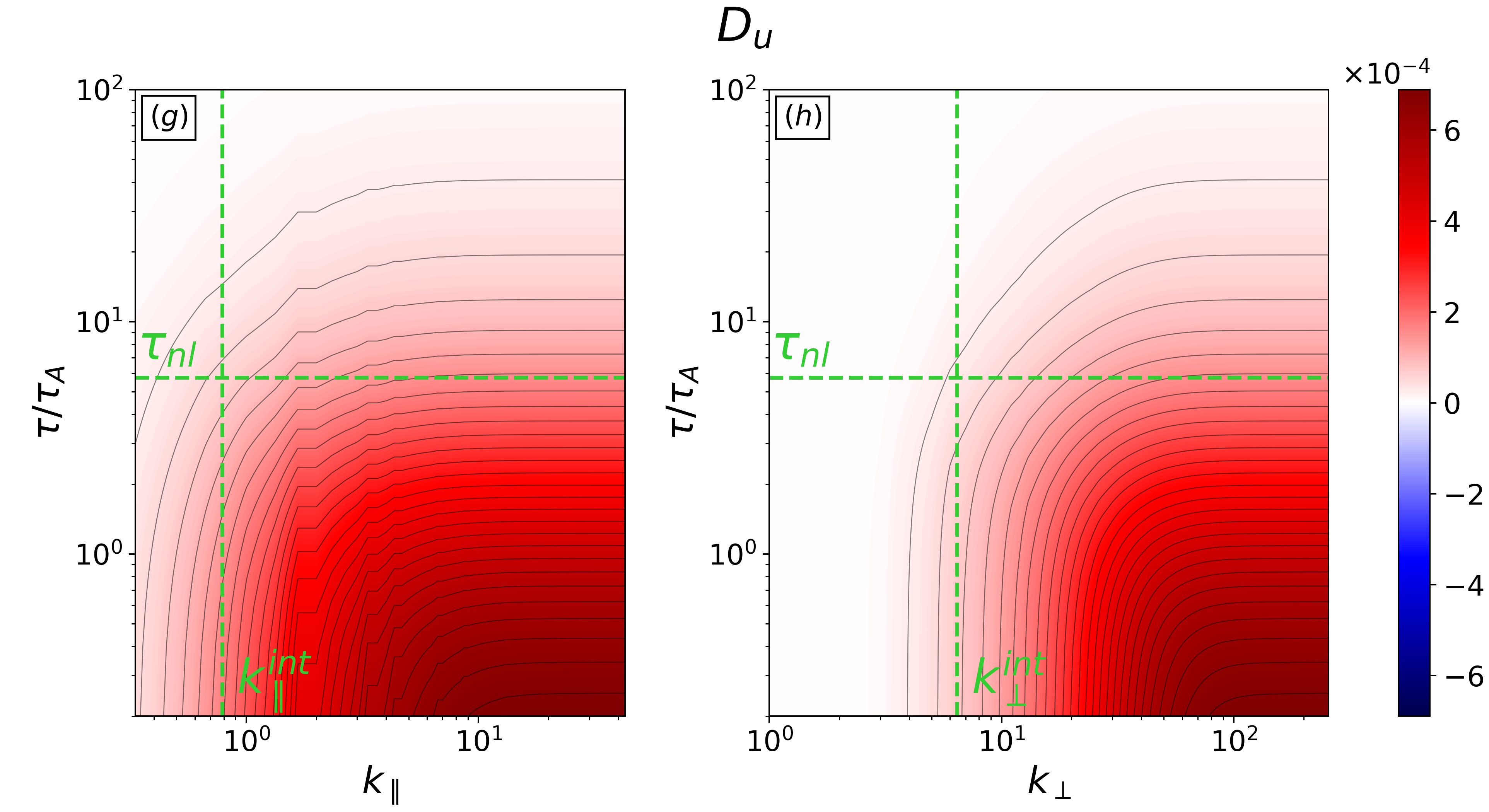}
}
\caption{$(k_{\parallel},\,\tau)$ and $(k_{\perp},\,\tau)$ projections of $I_{_B}$ (a)-(b), $I_{u}$ (c)-(d), $D_{_B}$ (e)-(f), and $D_{u}$ (g)-(h). Vertical green dashed lines indicate parallel and perpendicular integral scales $k_{\parallel}^{int}$ and $k_{\perp}^{int}$, while horizontal green dashed lines represent the nonlinear time $\tau_{nl}$.}
\label{I_D}
\end{figure*}
%%%%%%%%%%%%%%%%%%%%%%%%%%%%%%%%%%%%%%%%%%%%%%%%%%

This section provides a detailed derivation of the spatio-temporal coarse grained magnetohydrodynamic (MHD) equations employed in the Letter. 

We introduce a space-time low-pass filter defined as 
\begin{equation}
\overline{q}\bigl(\textbf{x},\,t,\,\textbf{k},\,\tau\bigr)\!=\! \sum\limits_{\textbf{k}^{\prime}<\textbf{k}}\int dt^{\prime} \,Q\bigl(\textbf{k}^{\prime},\,t^{\prime}\bigr)\,e^{i\,\textbf{k}^{\prime}\cdot\textbf{x}}\,G_{\tau}\bigl(t\!-\!t^{\prime}\bigr),
\label{Filter}
\end{equation}
where $q(\textbf{x},\,t)$ is a generic quantity with spatial Fourier transform $Q(\textbf{k},\,t)$, and $G_{\tau}$ is a boxcar function with width $\tau$. The filtered quantity $\overline{q}$ contains fluctuations with wavenumbers $<\textbf{k}$ and time scales $>\tau$. The corresponding density-weighted filter is $\widehat{q}\!=\!\overline{\rho\,q}/\overline{\rho}$. We consider the MHD equations
\begin{equation}
\begin{gathered}
\partial_t \rho + \nabla \!\cdot\! \bigl( \rho \textbf{u} \bigr) \!= 0,
\\[5pt]
\partial_t \bigl( \rho \textbf{u} \bigr) + \nabla \!\cdot\!\bigl( \rho \textbf{u} \textbf{u} \bigr) \!= -\nabla P + \textbf{J} \times \textbf{B} + \nabla \!\cdot\! \boldsymbol{\Pi} + \textbf{F}_u,
\\[5pt]
\partial_t \textbf{B} = \!\nabla \!\times\! \bigl( \textbf{u} \times \textbf{B} -\eta\,\textbf{J} \bigr) + \textbf{F}_{_B}, 
\end{gathered}
\label{Full_MHD}
\end{equation}
with $\rho$, $\textbf{u}$ and $P$ indicating the plasma density, velocity and pressure. $\textbf{B}$ is the magnetic field, $\textbf{J}\!=\!\nabla\!\times\!\textbf{B}$, $\eta$ is the magnetic diffusivity. $\boldsymbol{\Pi} \!=\! \rho\,\nu \bigl[ \nabla\textbf{u} + \!\nabla\textbf{u}^T\! - \!\bigl( 2/3 \bigr) \bigl(\nabla\!\cdot\!\textbf{u}\bigr) \textbf{I} \bigr]$ is the viscous stress tensor, with $\nu$ being the viscosity, $\textbf{I}$ is the identity matrix, and $T$ indicates the transpose operation. $\textbf{F}_u$ and $\textbf{F}_{_B}$ are large scale turbulent forcing terms. 

By applying the low-pass filter in Eq.~\ref{Filter} to Eqs.~\ref{Full_MHD}, we obtain
\begin{equation}
\begin{gathered}
\partial_t \overline{\rho} + \nabla \!\cdot\! \bigl( \overline{\rho} \, \widehat{\textbf{u}} \bigr) \!= 0,
\\[5pt]
\partial_t \bigl( \overline{\rho} \, \widehat{\textbf{u}} \bigr) + \nabla \!\cdot\!\bigl( \overline{\rho} \, \widehat{\textbf{u} \textbf{u}} \bigr) \!= -\nabla \overline{P} + \overline{\textbf{J} \times \textbf{B}} + \nabla \!\cdot\! \overline{\boldsymbol{\Pi}} + \overline{\textbf{F}}_u,
\\[5pt]
\partial_t \overline{\textbf{B}} = \!\nabla \!\times\! \bigl( \overline{\textbf{u} \times \textbf{B}} \bigr) - \eta\, \nabla \!\times\! \overline{\textbf{J}} + \overline{\textbf{F}}_{_B}.
\end{gathered}
\label{Filtered_MHD_01}
\end{equation}
The strategy now is expressing Eqs.~\ref{Filtered_MHD_01} in terms of the low-pass filtered quantities $\overline{\rho}$, $\widehat{\textbf{u}}$ and $\overline{\textbf{B}}$, plus some subscale terms. While the continuity equation is already in the correct form, the momentum and induction equations need some simple algebraic manipulation. To this end, we introduce
\begin{equation}
\begin{gathered}
\widehat{\textbf{u} \textbf{u}} = \widehat{\textbf{u}} \widehat{\textbf{u}} + \bigl( \widehat{\textbf{u} \textbf{u}} - \widehat{\textbf{u}} \widehat{\textbf{u}} \bigr) = \widehat{\textbf{u}} \widehat{\textbf{u}} + \boldsymbol{\tau}_u,
\\[5pt]
\overline{\textbf{J}\times\textbf{B}} = \overline{\textbf{J}} \times \overline{\textbf{B}} + \bigl( \overline{\textbf{J}\times\textbf{B}} - \overline{\textbf{J}} \times \overline{\textbf{B}} \bigr) = \overline{\textbf{J}} \times \overline{\textbf{B}} + \boldsymbol{\tau}_{_B},
\\[5pt]
\overline{\textbf{u}\times\textbf{B}} = \widehat{\textbf{u}} \times \overline{\textbf{B}} + \bigl( \overline{\textbf{u}\times\textbf{B}} - \widehat{\textbf{u}} \times \overline{\textbf{B}} \bigr) = \widehat{\textbf{u}} \times \overline{\textbf{B}} - \boldsymbol{\tau}_{_E},
\end{gathered}
\label{Subscale}
\end{equation}
where $\boldsymbol{\tau}_u$, $\boldsymbol{\tau}_{_B}$ and $\boldsymbol{\tau}_{_E}$ are the subscale stress tensor, Lorentz force, and electric field. The momentum equation thus becomes
\begin{equation}
\begin{gathered}
\partial_t \bigl( \overline{\rho} \, \widehat{\textbf{u}} \bigr) + \nabla \!\cdot\!\bigl( \overline{\rho} \, \widehat{\textbf{u}} \widehat{\textbf{u}} \bigr) \!= -\nabla \overline{P} + \overline{\textbf{J}} \times \overline{\textbf{B}} + 
\\[5pt]
- \nabla \!\cdot\! \bigl( \overline{\rho} \, \boldsymbol{\tau}_u \bigr) + \boldsymbol{\tau}_{_B} + \nabla \!\cdot\! \overline{\boldsymbol{\Pi}} + \overline{\textbf{F}}_u,
\end{gathered}
\label{Momentum}
\end{equation}
while the induction equation becomes
\begin{equation}
\begin{gathered}
\partial_t \overline{\textbf{B}} = \!\nabla \!\times\! \bigl( \widehat{\textbf{u}} \times \overline{\textbf{B}} \bigr) - \nabla \!\times\! \boldsymbol{\tau}_{_E} - \eta\, \nabla \!\times\! \overline{\textbf{J}} + \overline{\textbf{F}}_{_B}.
\end{gathered}
\label{Induction}
\end{equation}
Equations for the low-pass filtered kinetic and magnetic energies are obtained by taking the dot product of Eq.~\ref{Momentum} with $\widehat{\textbf{u}}$, and of Eq.~\ref{Induction} with $\overline{\textbf{B}}$, which gives
\begin{equation}
\begin{gathered}
\partial_t \left( \frac{1}{2} \overline{\rho} \widehat{u}^2 \right) + \nabla \!\cdot\! \left( \frac{1}{2} \overline{\rho} \widehat{u}^2 \widehat{\textbf{u}}  \right) = - \nabla \overline{P} \!\cdot\! \widehat{\textbf{u}} +
\\[5pt]
- \bigl( \widehat{\textbf{u}} \times \overline{\textbf{B}} \bigr) \!\cdot\!\overline{\textbf{J}} - \bigl[ \nabla\!\cdot\! \bigl( \overline{\rho} \, \boldsymbol{\tau}_u \bigr) \bigr] \!\cdot\!\widehat{\textbf{u}} + \boldsymbol{\tau}_{_B} \!\cdot\! \widehat{\textbf{u}} +
\\[5pt]
+ \bigl( \nabla \!\cdot\! \overline{\boldsymbol{\Pi}} \bigr) \!\cdot\! \widehat{\textbf{u}} + \overline{\textbf{F}}_u \!\cdot\! \widehat{\textbf{u}},
\\[5pt]
\partial_t \left( \frac{1}{2} \overline{B}^2 \right) - \nabla \!\cdot\! \bigl[ \bigl( \widehat{\textbf{u}} \times \overline{\textbf{B}} \bigr) \times \overline{\textbf{B}} \bigr] = \bigl( \widehat{\textbf{u}} \times \overline{\textbf{B}} \bigr) \!\cdot\!\overline{\textbf{J}} +
\\[5pt]
- \bigl( \nabla \times \boldsymbol{\tau}_{_E} \bigr) \!\cdot\! \overline{\textbf{B}} - \eta\, \bigl( \nabla \!\times\! \overline{\textbf{J}} \bigr) \!\cdot\! \overline{\textbf{B}} + \overline{\textbf{F}}_{_B} \!\cdot\! \overline{\textbf{B}}.
\end{gathered}    
\label{Energy_MHD}
\end{equation}
The global energy balance is finally obtained by averaging Eqs.~\ref{Energy_MHD} over the system size, which gives
\begin{equation}
\begin{gathered}
\partial_t \biggl< \frac{1}{2}\,\overline{\rho}\,\widehat{u}^2 \biggr> + J_u = -W_{_P} -\Pi_u +W_{e.m.} +
\\[5pt]
-D_u +I_u,
\\[5pt]
\partial_t \biggl< \frac{1}{2}\,\overline{B}^2 \biggr> + J_{_B} = -\Pi_{_B} -W_{e.m.} -D_{_B} +I_{_B},    
\end{gathered}    
\label{Global_MHD}
\end{equation}
where $\langle\cdot\rangle$ indicates the spatial average, and energy transfer channels are
\begin{equation}
\begin{gathered}
J_u\!=\! \biggl< \nabla \!\cdot\! \left( \frac{1}{2} \overline{\rho} \widehat{u}^2 \widehat{\textbf{u}}  \right) \biggr>,
\quad 
W_{_P}\!=\!\bigl< \nabla \overline{P} \!\cdot\! \widehat{\textbf{u}} \bigr>,
\\[7pt]
\Pi_u\!=\! \bigl< \bigl[ \nabla\!\cdot\! \bigl( \overline{\rho} \, \boldsymbol{\tau}_u \bigr) \bigr] \!\cdot\!\widehat{\textbf{u}} - \boldsymbol{\tau}_{_B} \!\cdot \widehat{\textbf{u}} \bigr>, 
\\[7pt]
I_u\!=\!\bigl< \overline{\textbf{F}}_u \!\cdot \widehat{\textbf{u}} \bigr>,
\quad
D_u\!=\!\bigl< - \bigl( \nabla \!\cdot\! \overline{\boldsymbol{\Pi}} \bigr) \!\cdot\! \widehat{\textbf{u}} \bigr>,
\\[7pt]
J_{_B}\!=\!\biggl< - \nabla \!\cdot\! \bigl[ \bigl( \widehat{\textbf{u}} \times \overline{\textbf{B}} \bigr) \times \overline{\textbf{B}} \bigr] \biggr>,
\\[7pt]
W_{e.m.}\!=\!\bigl< -\bigl( \widehat{\textbf{u}}\times\overline{\textbf{B}} \bigr) \!\cdot\! \overline{\textbf{J}} \bigr>, 
\quad
\Pi_{_B}\!=\!\bigl< \bigl( \nabla \times \boldsymbol{\tau}_{_E} \bigr) \!\cdot\! \overline{\textbf{B}} \bigr>,
\\[7pt]
I_{_B}\!=\!\bigl< \overline{\textbf{F}}_{_B} \!\cdot \overline{\textbf{B}} \bigr>,
\quad
D_{_B}\!=\!\bigl< \eta\, \bigl( \nabla \times \overline{\textbf{J}} \bigr) \!\cdot\! \overline{\textbf{B}} \bigr>.
\end{gathered}
\label{Full_ETC}
\end{equation}
If we assume no energy transport across the system boundaries, or if the system is periodic, the kinetic and magnetic energy fluxes $J_u$ and $J_{_B}$ vanish, while other terms can be rewritten as
\begin{equation}
\begin{gathered}
W_{_P} \!=\! \bigl< -\overline{P}\, \nabla\!\cdot\!\widehat{\textbf{u}} \bigr>,
\quad 
D_u\!=\! \bigl< \overline{\boldsymbol{\Pi}} : \!\nabla\widehat{\textbf{u}} \bigr>,
\\[7pt]
\Pi_u \!=\! \bigl< - \overline{\rho}\, \boldsymbol{\tau}_u \!:\! \nabla\widehat{\textbf{u}} - \boldsymbol{\tau}_{_B} \!\cdot \widehat{\textbf{u}} \bigr>, 
\\[7pt]
\Pi_{_B}\!=\!\bigl< \boldsymbol{\tau}_{_E} \!\cdot \overline{\textbf{J}} \bigr>,
\quad
D_{_B}\!=\!\bigl< \eta\, \overline{\textbf{J}}^2 \bigr>.
\end{gathered}
\end{equation}

\section{Magnetic and kinetic injection and dissipation rates}

In this section, we show the energy injection rate and the energy dissipation rate decomposed into their magnetic and kinetic components, for the $(\beta\!=\!0.5,\,\sigma_{_C}\!\simeq\!0)$ simulation analyzed in the Letter.

The first row of Fig.~\ref{I_D} shows the magnetic energy injection rate $I_{_B}$, panels (a)-(b), and the kinetic energy injection rate $I_u$, panels (c)-(d). Both terms exhibit a similar $(k_{\perp},\,k_{\parallel},\,\tau)$ dependence, with comparable amplitudes. In particular, both $I_{_B}$ and $I_u$ are positive and increase with wavenumber, up to $k_{\parallel}\!\simeq\!1.3$ and $k_{\perp}\!\simeq\!6$, after which they saturates as there is no contribution stemming from larger wavenumbers. $I_{_B}$ and $I_u$ increase from $\tau\!\simeq\!15$ to $\tau\!\simeq\!1.5$, being almost constant elsewhere. The fact that $I_{_B}$ and $I_u$ have analogous features, including a broadband frequency response, is a consequence of the design of our turbulent drivers $\textbf{F}_{_B}$ and $\textbf{F}_u$, both of them injecting magnetic and velocity fluctuations in the same range of wavenumbers and frequencies, with similar amplitudes. 

The energy dissipation rate components are shown in the second row of Fig.~\ref{I_D}, where $D_{_B}$ is the magnetic energy dissipation rate, panels (e)-(f), and $D_u$ is the kinetic energy dissipation rate, panels (g)-(h). Both $D_{_B}$ and $D_{u}$ are weak at large scales, and quickly increase toward large wavenumbers and small $\tau$. This behavior implies that dissipation affects mainly large wavenumbers and high frequencies, while low $\omega$ fluctuations are not dissipated. The similarities observed for $D_{_B}$ and $D_{u}$ are a consequence of the fact that we have chosen both the viscosity $\nu$ and magnetic diffusivity $\eta$ to have the same value. 

\section{Imbalanced turbulence run and high-$\beta$ run}

%%%%%%%%%%%%%%%%%%%%%%%%%%%%%%%%%%%%%%%%%%%%%%%%%%
\begin{figure*}[t]
\centering
\subfloat{
\includegraphics[width=0.32\linewidth]{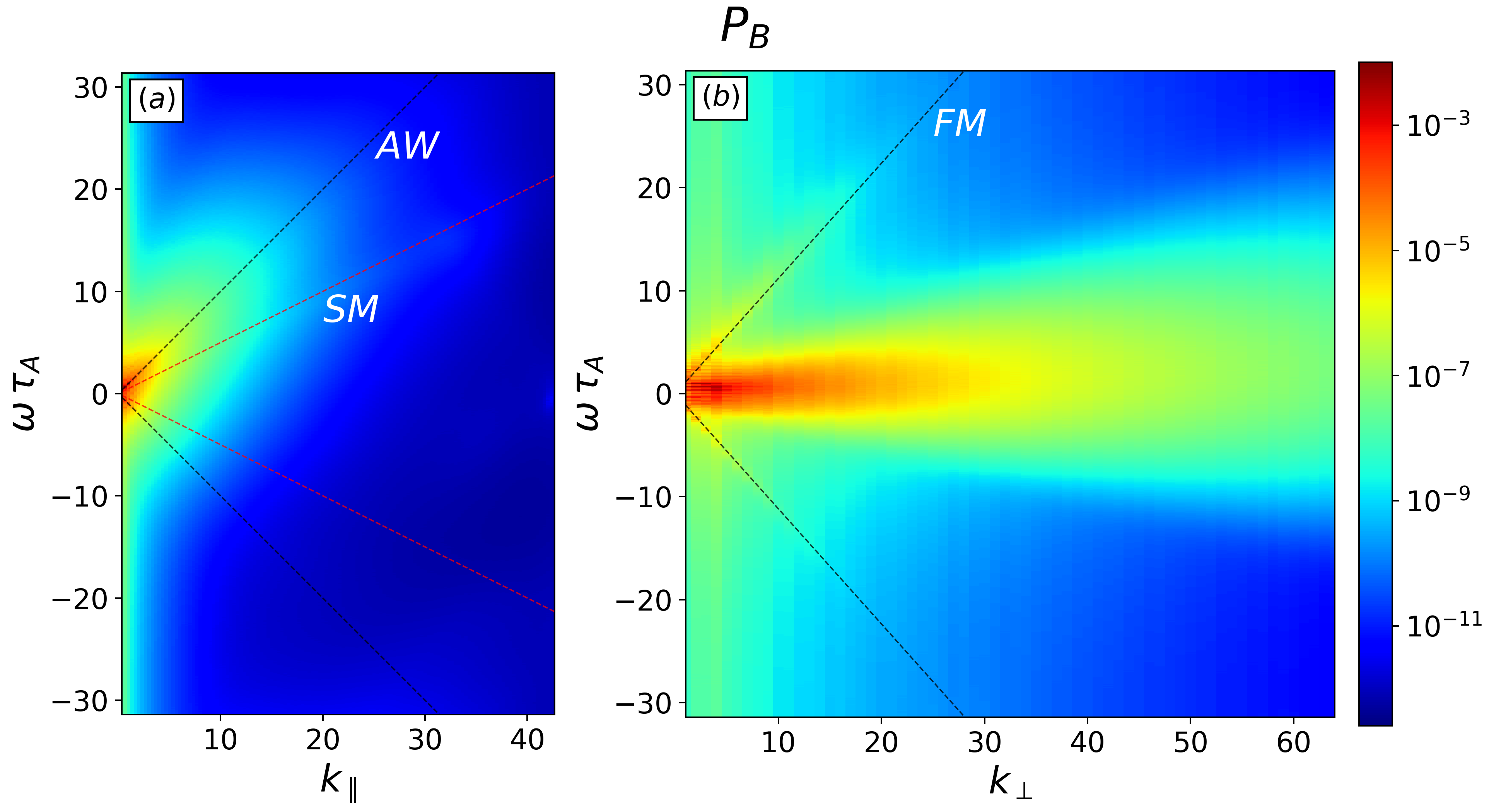}
}
\subfloat{
\includegraphics[width=0.32\linewidth]{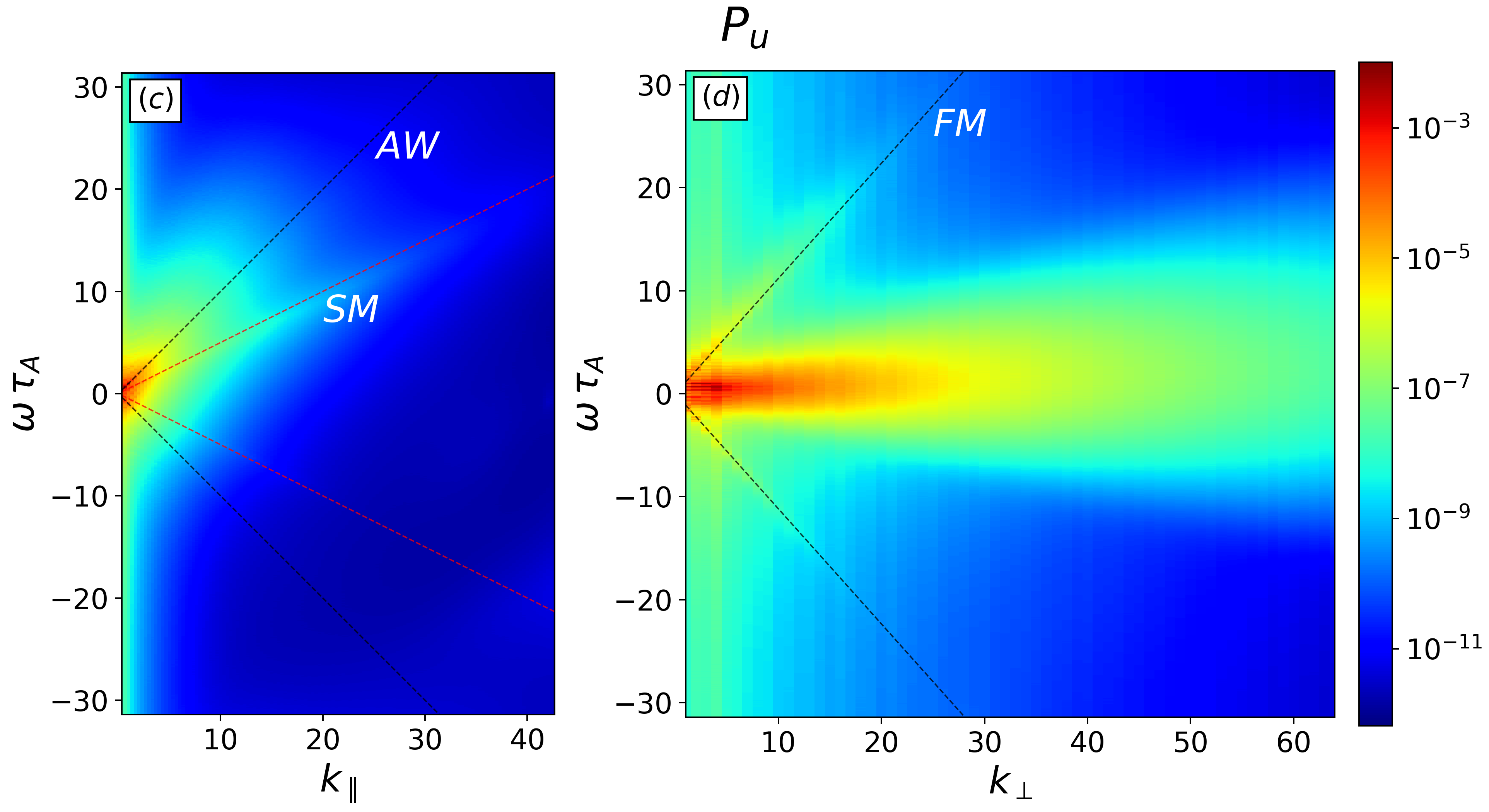}
}
\subfloat{
\includegraphics[width=0.32\linewidth]{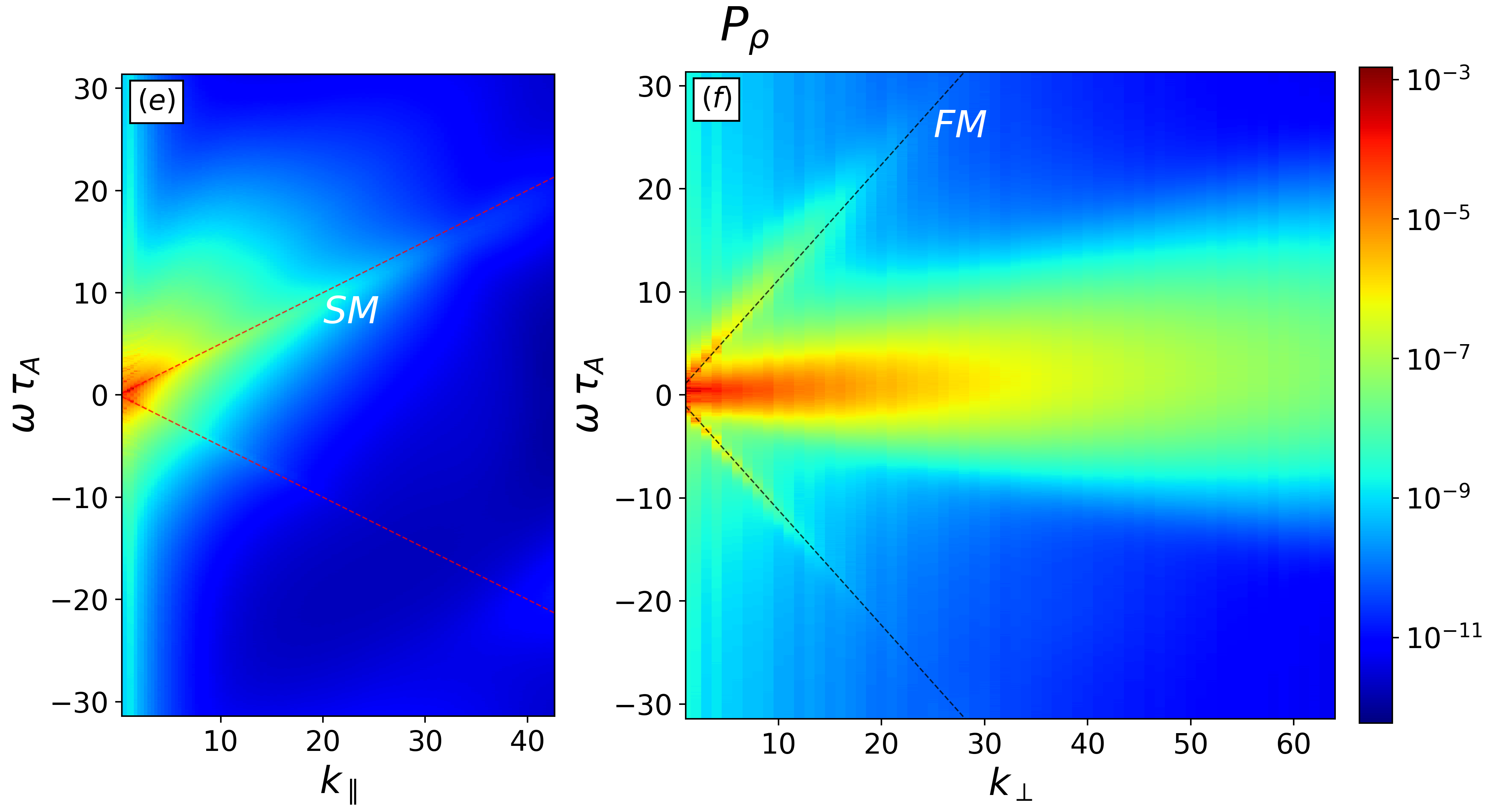}
}
\\
\subfloat{
\includegraphics[width=0.32\linewidth]{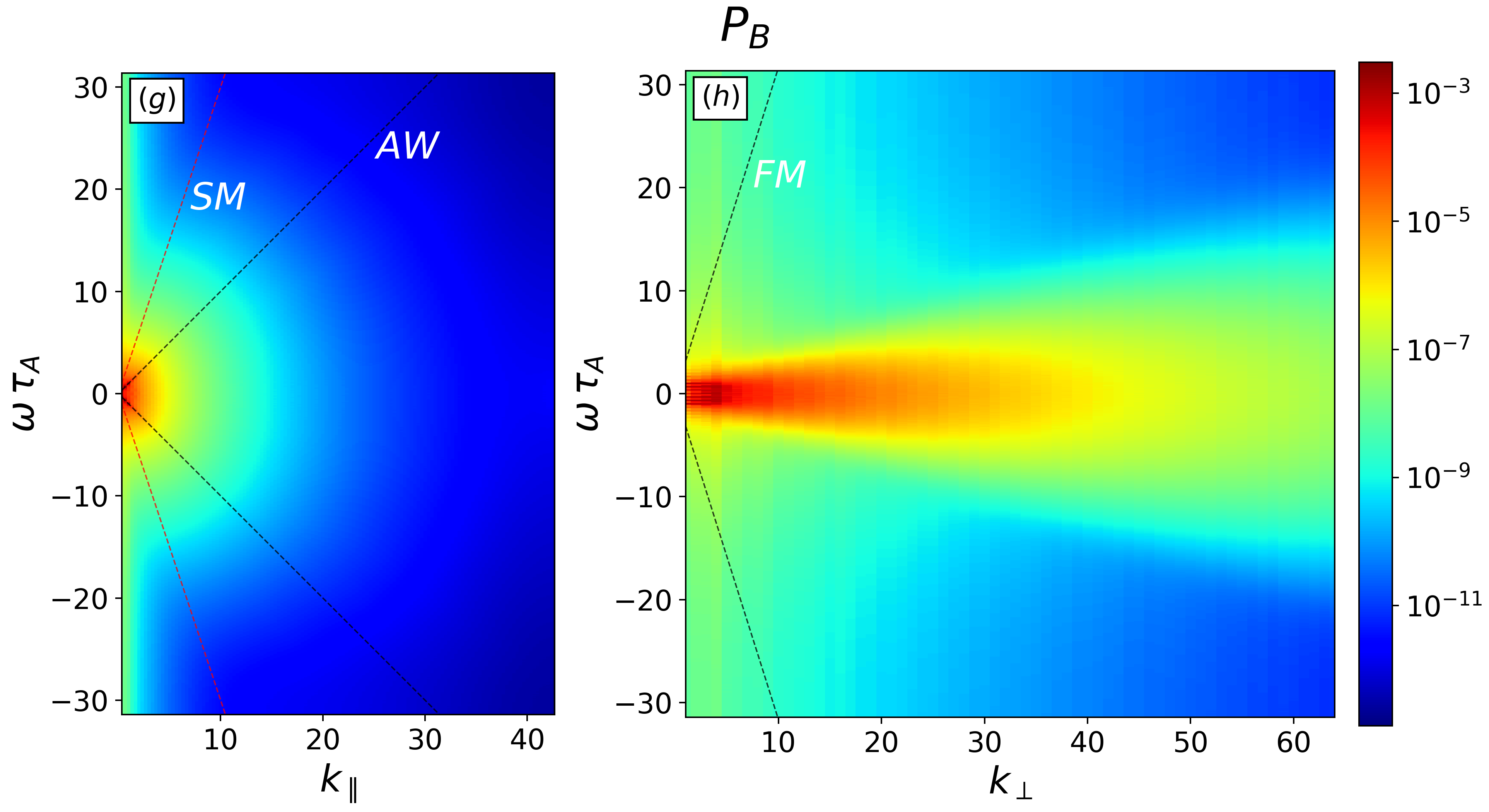}
}
\subfloat{
\includegraphics[width=0.32\linewidth]{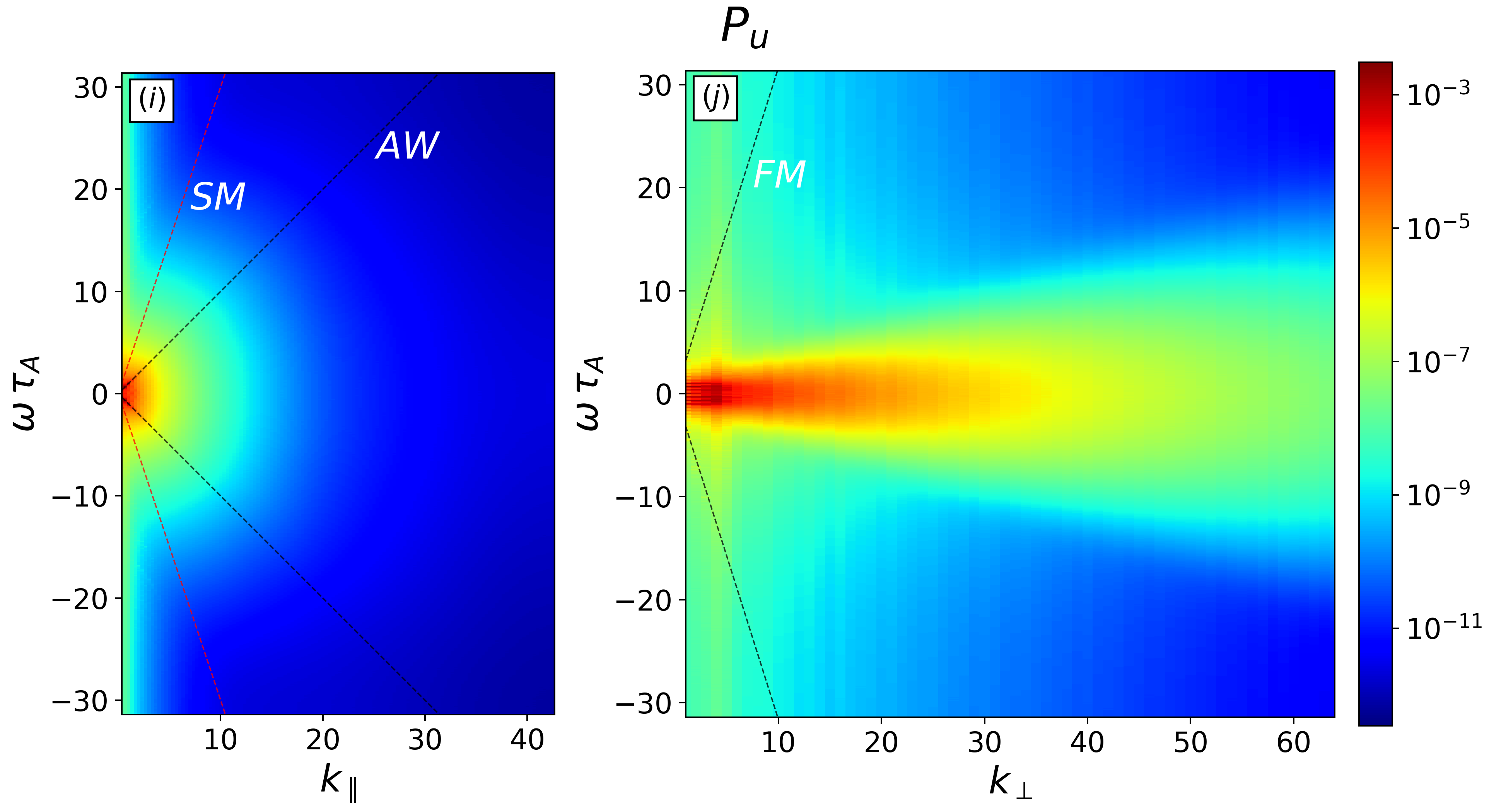}
}
\subfloat{
\includegraphics[width=0.32\linewidth]{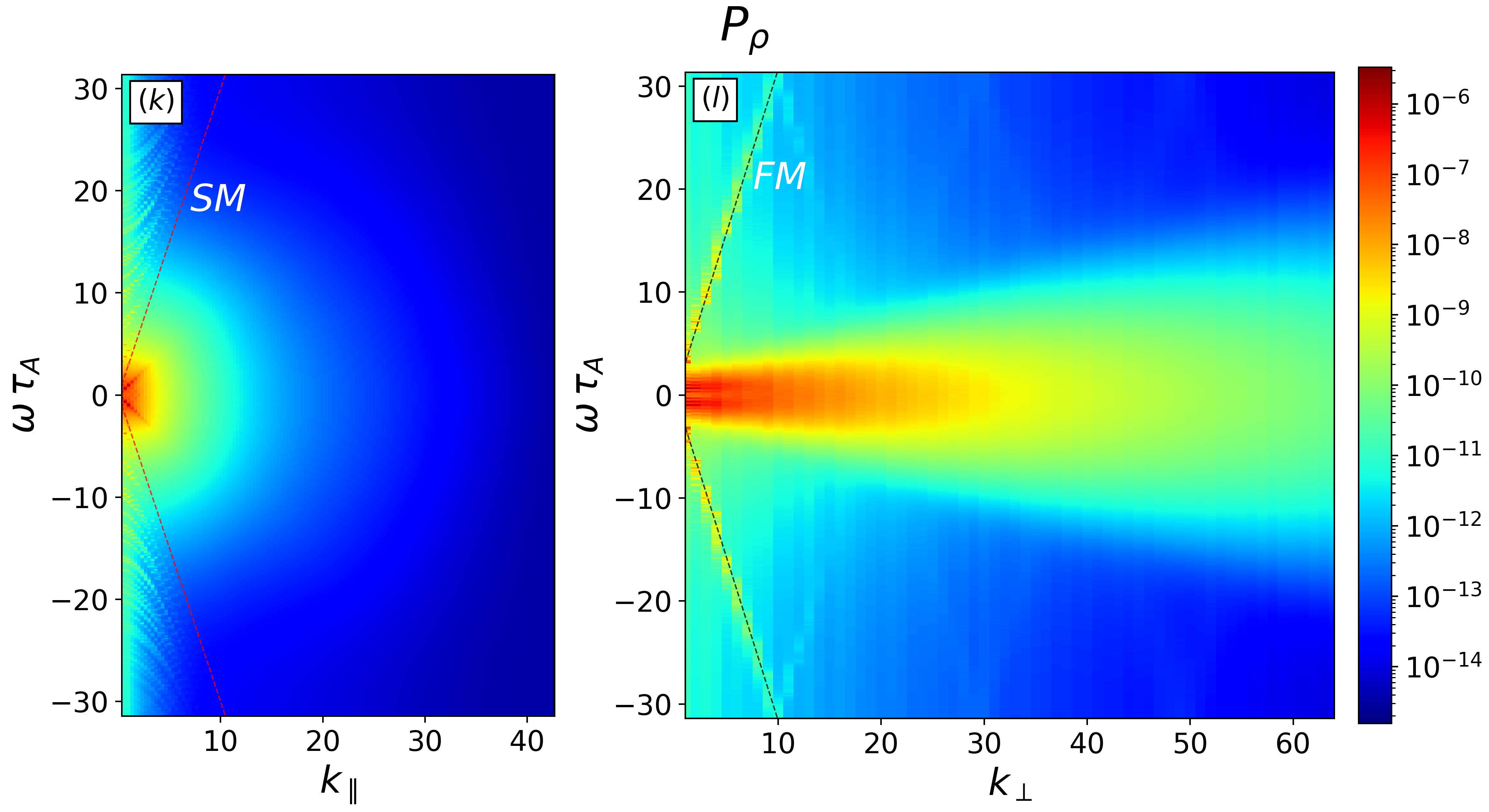}
}
\caption{$(k_{\parallel},\,\omega)$ and $(k_{\perp},\,\omega)$ projections of magnetic field, velocity and density spectra $P_{_B}$, $P_{u}$ and $P_{\rho}$ for the imbalanced turbulence run (a)-(f), and for the high-$\beta$ turbulence run (g)-(l). Dashed lines indicate the the dispersion relations of Aflvén wave (AW) and slow modes (SM) for parallel propagation ($k_{\perp}\!=\!0$), and of fast modes (FM) for perpendicular propagation ($k_{\parallel}\!=\!0$).}
\label{P_ibal_beta}
\end{figure*}
%%%%%%%%%%%%%%%%%%%%%%%%%%%%%%%%%%%%%%%%%%%%%%%%%%

In this section, we show results from our spatio-temporal coarse graining (CG) analysis applied to other two 3D magnetohydrodynamic (MHD) simulations of turbulence, realized using the code \textit{Athena++}. Both simulations are initialized with exactly the same parameters as the simulation analyzed in the Letter, but we vary the cross helicity and plasma $\beta$. Cross helicity $\sigma_{_C}$ is defined as 
\begin{equation}
\sigma_{_C} = \frac{2 \, \sqrt{\rho} \, \textbf{u} \cdot \delta\textbf{B}}{\rho u^2 + \delta B^2},    
\end{equation}
where $\delta \textbf{B} \!=\! \textbf{B} - \textbf{B}_0$ (with $\textbf{B}_0\!=\!B_0\,\widehat{\textbf{z}}$ being the guide field). The first simulation (run A) has $(\beta\!=\!0.5,\,\sigma_{_C}\!\simeq\!0.73)$, representing near-Sun solar wind turbulence, where $\sigma_{_C}$ reaches large values \citep{chen2020evolution}. The second simulation (run B) has $(\beta\!=\!18,\,\sigma_{_C}\!\simeq\!0)$, typical of solar wind turbulence in the outer heliosphere \citep{fraternale2022exploring}. Due to the different $\sigma_{_C}$ and $\beta$, both simulations reach a fully developed state where magnetic, velocity and density fluctuations exhibit root mean square (rms) amplitudes slightly different from the simulation analyzed in the Letter. Specifically, run A has $\delta B_{rms}/B_0\!\simeq\!0.17$ and $\delta u_{rms}/c_{_A}\!\simeq\!0.17$ (with $c_{_A}$ being the Alfvén speed), while run B has $\delta B_{rms}/B_0\!\simeq\!0.15$ and $\delta u_{rms}/c_{_A}\!\simeq\!0.14$. Density fluctuations reach an rms amplitude of $\delta \rho_{rms}/\rho_0\!\simeq\!0.06$ (with $\rho_0$ being the initial density) in run A, while run B has $\delta \rho_{rms}/\rho_0\!\simeq\!0.003$, due to high $\beta$ suppressing compressible fluctuations \citep{biskamp1997nonlinear}. 

%%%%%%%%%%%%%%%%%%%%%%%%%%%%%%%%%%%%%%%%%%%%%%%%%%
\begin{figure*}[t]
\centering
\subfloat{
\includegraphics[width=0.32\linewidth]{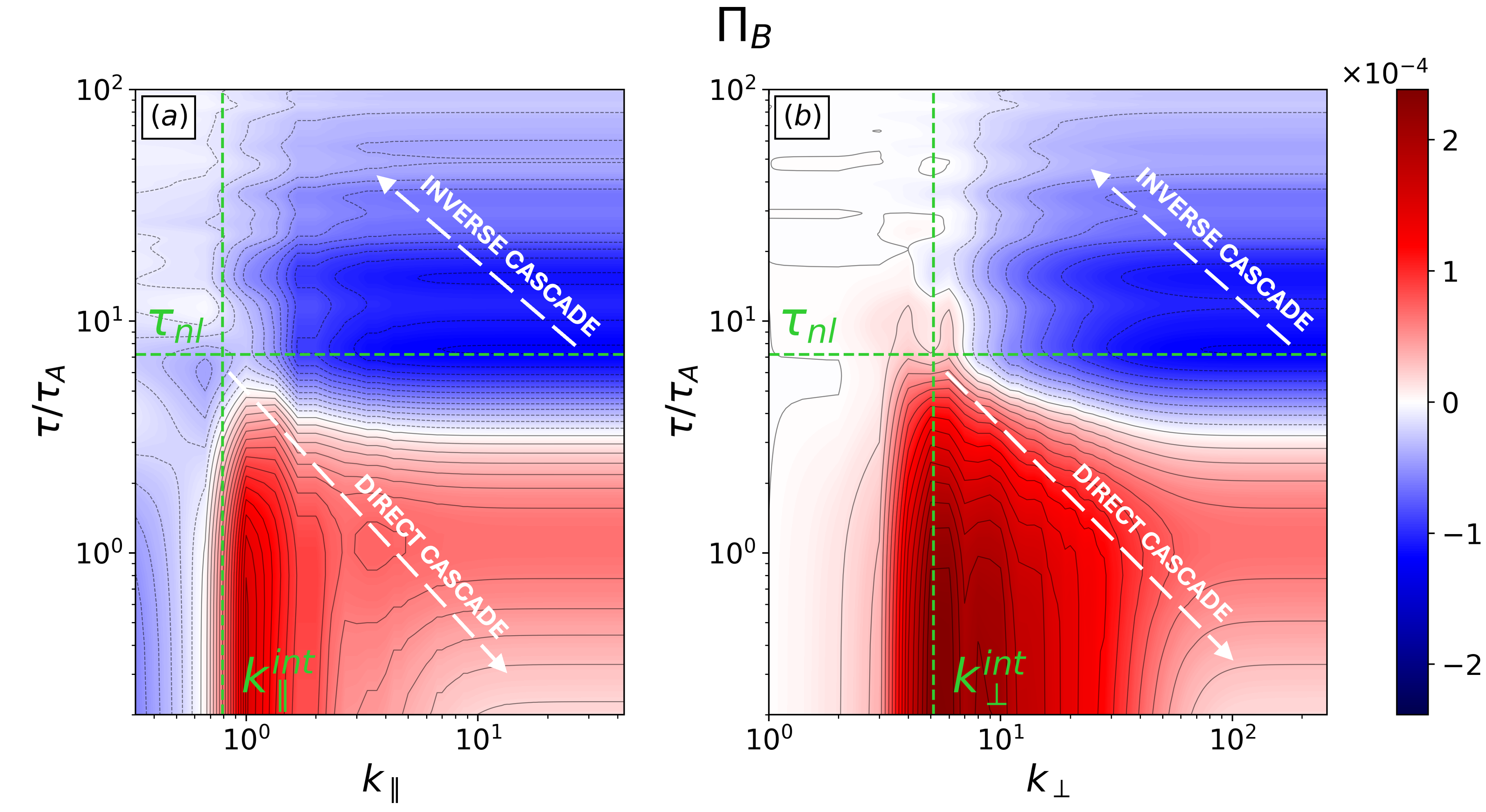}
}
\subfloat{
\includegraphics[width=0.32\linewidth]{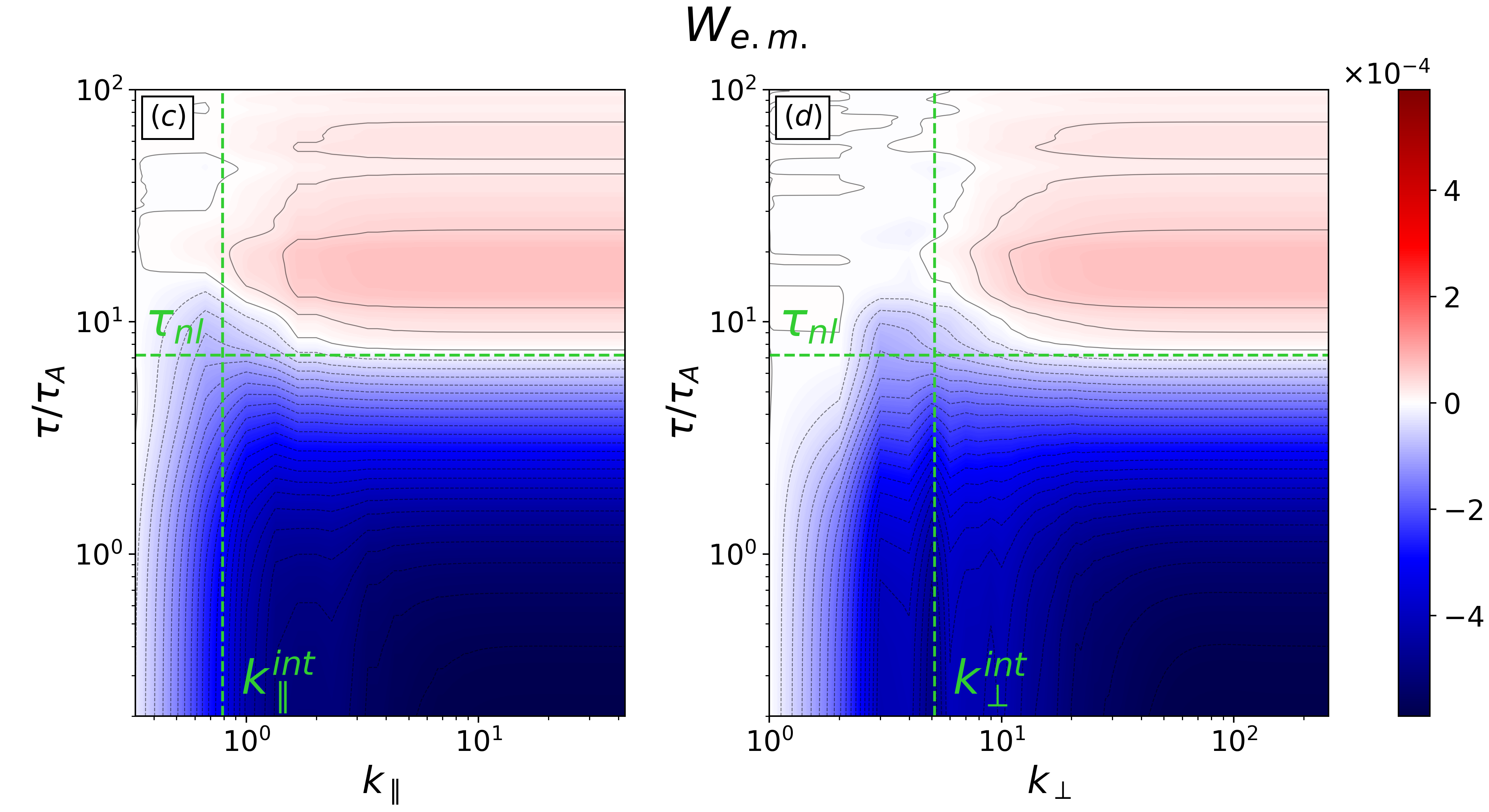}
}
\subfloat{
\includegraphics[width=0.32\linewidth]{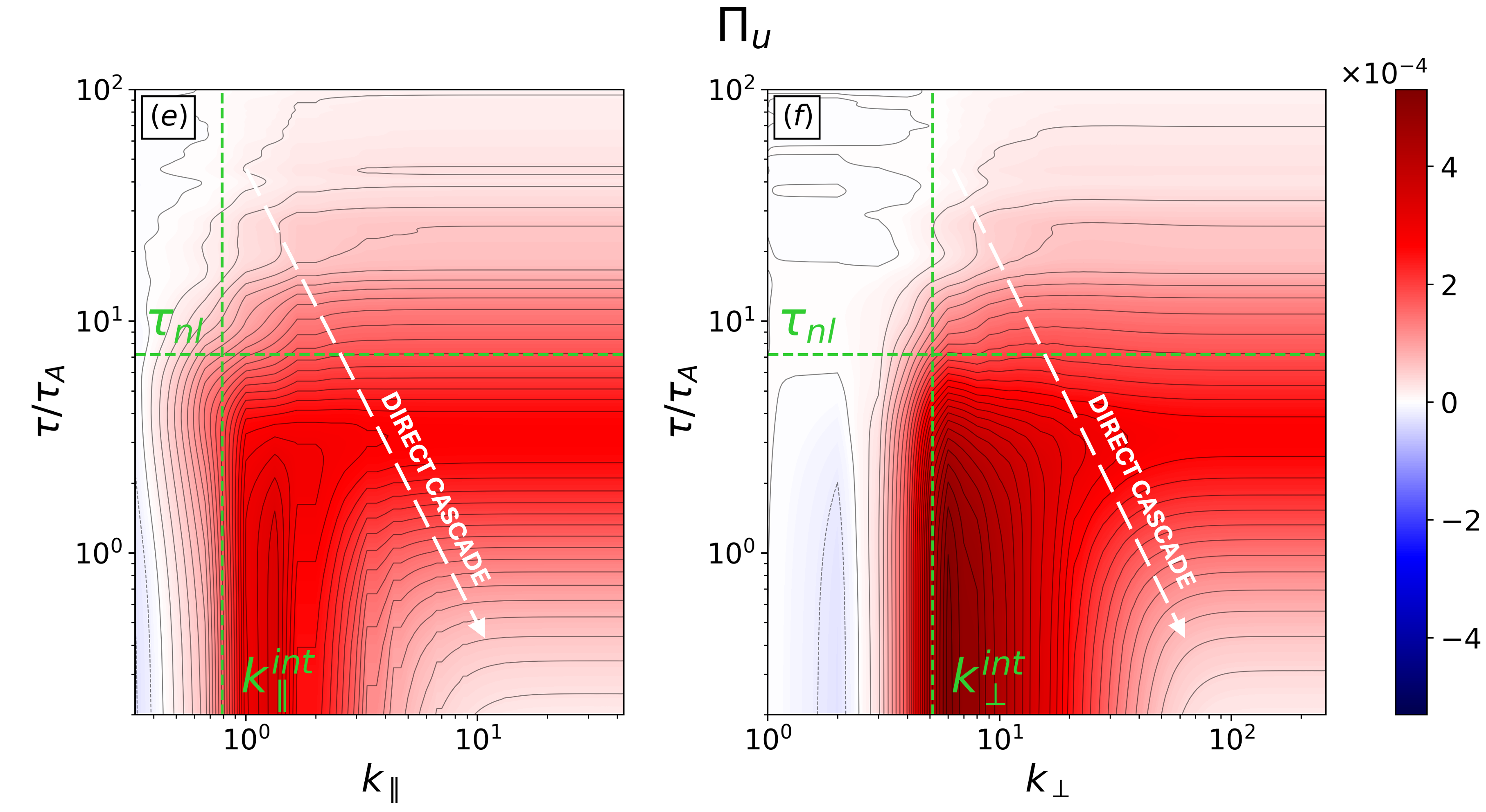}
}
\\
\subfloat{
\includegraphics[width=0.32\linewidth]{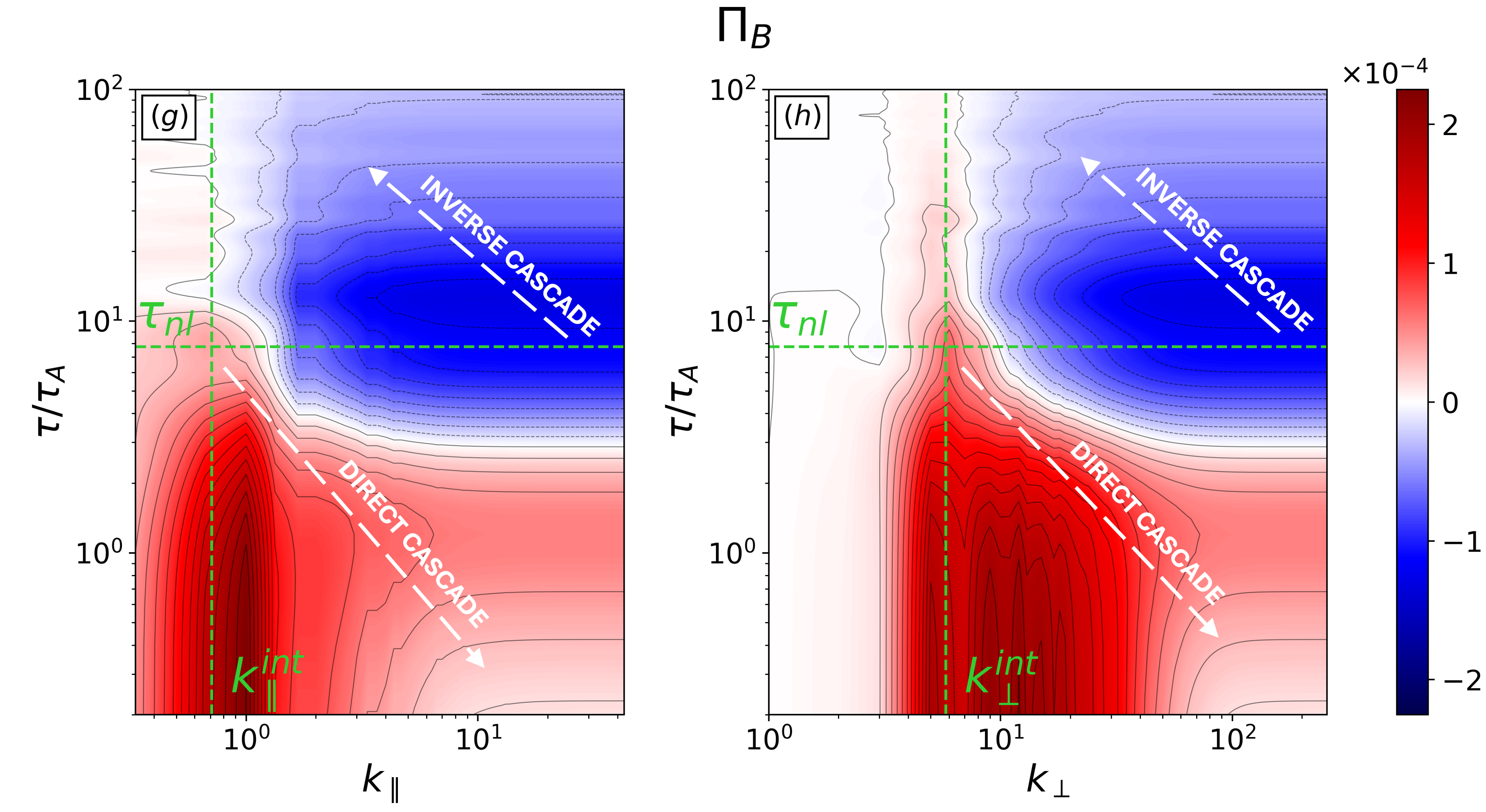}
}
\subfloat{
\includegraphics[width=0.32\linewidth]{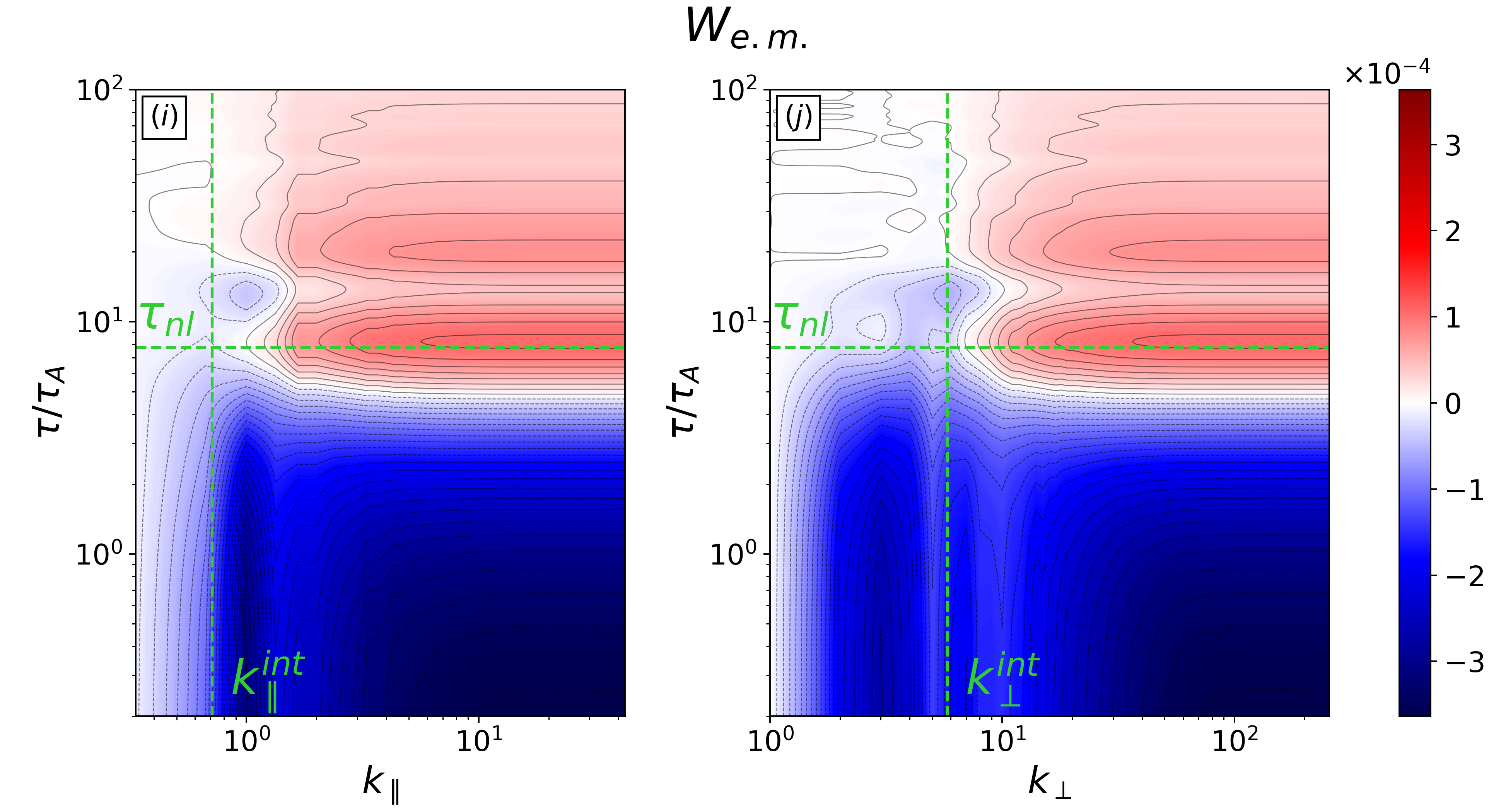}
}
\subfloat{
\includegraphics[width=0.32\linewidth]{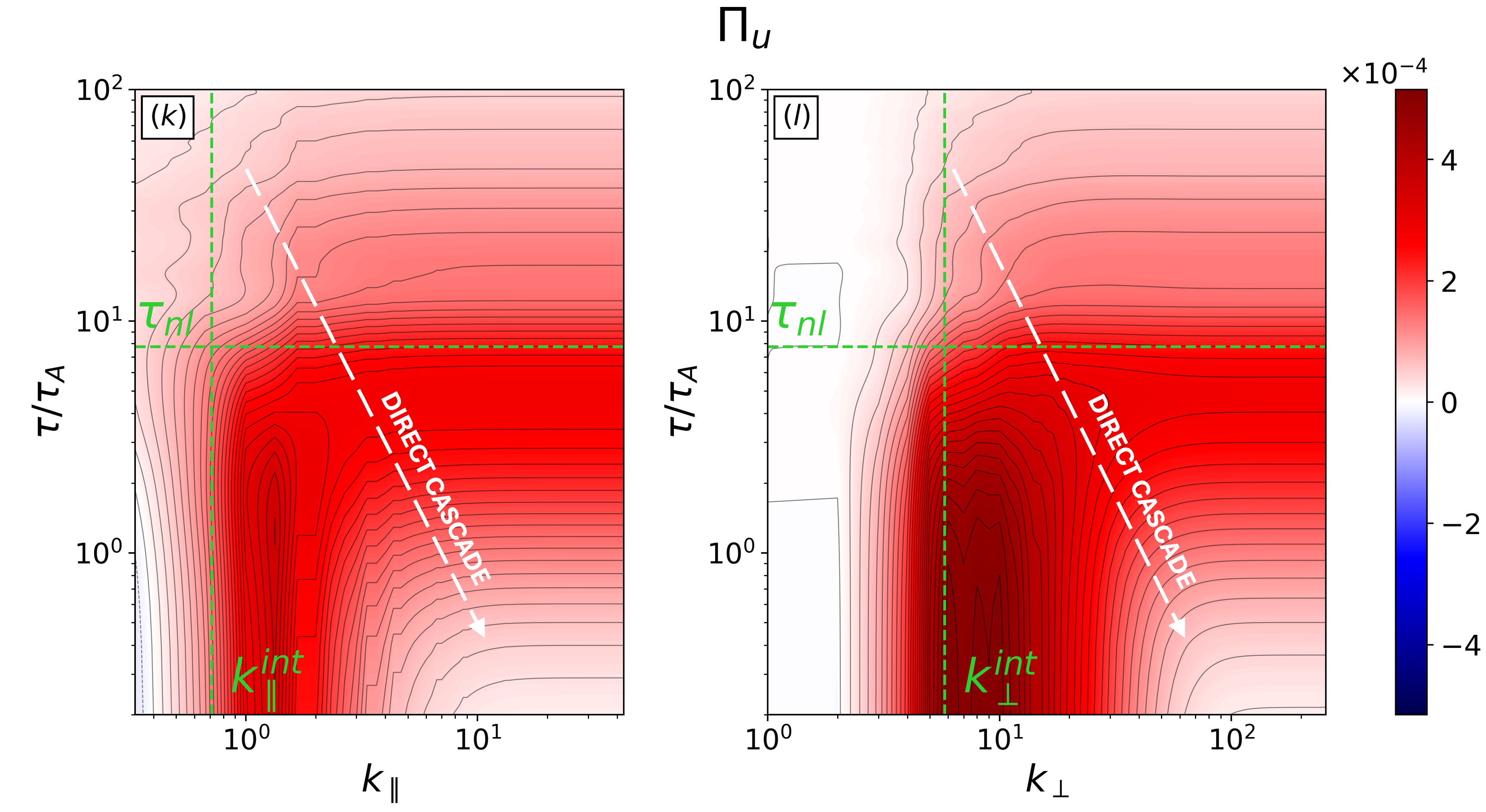}
}
\caption{$(k_{\parallel},\,\tau)$ and $(k_{\perp},\,\tau)$ projections of $\Pi_{_B}$, $W_{e.m.}$ and $\Pi_{u}$ for the imbalanced turbulence run (a)-(f), and for the high-$\beta$ turbulence run (g)-(l). Vertical green dashed lines indicate parallel and perpendicular integral scales $k_{\parallel}^{int}$ and $k_{\perp}^{int}$, while horizontal green dashed lines represent the nonlinear time $\tau_{nl}$.}
\label{ETC_ibal_beta}
\end{figure*}
%%%%%%%%%%%%%%%%%%%%%%%%%%%%%%%%%%%%%%%%%%%%%%%%%%

Figure~\ref{P_ibal_beta} shows $(k_{\parallel},\,\omega)$ and $(k_{\perp},\,\omega)$ projections of magnetic field, velocity and density spectra $P_{_B}$, $P_{u}$ and $P_{\rho}$, for run A, panels (a)-(f), and for run B, panels (g)-(l), with dashed lines indicating dispersion relations of Alfvén waves (AW), slow modes (SM), and fast modes (FM). We see that in both runs, most of magnetic and kinetic energy is concentrated at low $\omega$ and small $k_{\parallel}$, with a wider distribution in $k_{\perp}$, and only a negligible fraction of energy associated with waves. This is analogous to what we observe in the $(\beta\!=\!0.5,\,\sigma_{_C}\!\simeq\!0)$ simulation analyzed in the Letter. Density spectra exhibit a similar distribution as $P_{_B}$ and $P_u$ in $(k_{\parallel},\,k_{\perp},\,\omega)$ space, with $P_{\rho}$ being about three orders of magnitude smaller in run B with respect to run A, consistently with the lower level of compressibility expected for high $\beta$ turbulence. A regular wavy pattern is observed in the $(k_{\parallel},\,\omega)$ projection of $P_{\rho}$, at small $k_{\parallel}$ and toward high $\omega$, in run B. This pattern is caused by the finite energy contained in FMs, producing those features in $(k_{\parallel},\,\omega)$ space when averaged over $k_{\perp}$. A peculiar property of run A is that all spectra are skewed toward positive $\omega$ in $(k_{\parallel},\,\omega)$ space. This happens because high $\sigma_{_C}$ implies that modes propagating in the direction parallel to the guide field $\textbf{B}_0$ have less energy than modes propagating antiparallel to $\textbf{B}_0$ \citep{lugones2019spatio}. 

Figure~\ref{ETC_ibal_beta} shows $(k_{\parallel},\,\tau)$ and $(k_{\perp},\,\tau)$ projections of the magnetic energy cascade rate $\Pi_{_B}$, electromagnetic (e.m.) work $W_{e.m.}$, and kinetic energy cascade rate $\Pi_{u}$, for run A, panels (a)-(f), and for run B, panels (g)-(l), with black lines indicating their isocontours. As a reference, we show the parallel and perpendicular integral scales $k_{\parallel}^{int}$ and $k_{\perp}^{int}$ (horizontal green dashed lines), together with the nonlinear time $\tau_{nl}$ (vertical green dashed lines), as defined in the Letter. Since runs A and B are driven with the same method as the simulation analyzed in the Letter, with the same viscosity $\nu$ and magnetic diffusivity $\eta$, energy injection rates $I_{_B}$ and $I_{u}$, and energy dissipation rates $D_{_B}$ and $D_{u}$, exhibit the same properties as those of the $(\beta\!=\!0.5,\,\sigma_{_C}\!\simeq\!0)$ run, so we do not show them here, to avoid redundancy. The pressure work $W_{_P}$, quantifying compressible effects, is negligibly small with respect to the other energy transfer channels (ETCs) in both simulations (especially in run B). Hence, we do not show $W_{_P}$ for runs A and B here, since its contribution to the global energy balance is irrelevant. Here, we focus on analyzing $\Pi_{_B}$, $W_{e.m.}$ and $\Pi_{u}$, as their interplay is responsible for the origin of low $\omega$ fluctuations and for their contribution to the turbulent cascade. Figures~\ref{ETC_ibal_beta}(a)-(b) show that $\Pi_{_B}$ of run A has a bifurcation in $\tau$ space, being negative (inverse cascade) for $\tau\!\gtrsim\!\tau_{nl}$, and positive (direct cascade) for $\tau\!<\!\tau_{nl}$, where it peaks around integral scales. The same kind of behavior is observed for $\Pi_{_B}$ of run B, Fig.~\ref{ETC_ibal_beta}(g)-(h). Thus, consistently with the $(\beta\!=\!0.5,\,\sigma_{_C}\!\simeq\!0)$ run analyzed in the Letter, we find that the inverse magnetic energy cascade responsible for the formation of low $\omega$ (large $\tau$) magnetic fluctuations takes place in both imbalanced and high $\beta$ turbulence. In both simulations, magnetic energy flowing to large $\tau$ is converted into low $\omega$ kinetic energy by $W_{e.m.}$, which is positive for $\tau\!\gtrsim\!\tau_{nl}$, as seen in Fig.~\ref{ETC_ibal_beta}(c)-(d) for run A, and in Fig.~\ref{ETC_ibal_beta}(i)-(j) for run B. Finally, in both runs, low $\omega$ kinetic energy cascades to small spatial and temporal scales, as highlighted by $\Pi_u$, which is positive at all $(k,\,\tau)$ scales, with a stronger cascade rate around $k_{\parallel}^{int}$, $k_{\perp}^{int}$ and $\tau_{nl}$, as shown in Fig.~\ref{ETC_ibal_beta}(e)-(f) for run A, and in Fig.~\ref{ETC_ibal_beta}(k)-(l) for run B. Part of the kinetic energy transferred to small scales is then converted into magnetic energy by $W_{e.m.}$, being negative for $\tau\!<\!\tau_{nl}$. 

To summarize, we find that magnetic, velocity and density fluctuations in both imbalanced and high $\beta$ turbulence mainly reside in low $\omega$ modes, with wavenumbers almost perpendicular to the guide field. The ETCs analysis reveals that low $\omega$ magnetic modes are produced by an inverse magnetic energy cascade affecting fluctuations with time scales $\tau\!\gtrsim\!\tau_{nl}$. Low $\omega$ magnetic modes transfer part of their energy to low $\omega$ velocity fluctuations, explaining the abundance of kinetic energy at low $\omega$ in energy spectra. Low $\omega$ velocity fluctuations ultimately undergo a direct cascade toward higher wavenumbers and frequencies, meaning that low $\omega$ modes actually contribute to driving the turbulent cascade. This dynamics is consistent with what we find for the $(\beta\!=\!0.5,\,\sigma_{_C}\!\simeq\!0)$ simulation. 

Our analysis shows that the inverse low frequency magnetic field energy cascade, which is the main finding uncovered by our spatio-temporal CG method, appears to be a robust feature of plasma turbulence, observed in both balanced and imbalanced turbulence, and for both low and high $\beta$. We finally note that in both run A and run B, $W_{e.m.}$ at large $\tau$ is weaker in amplitude than $W_{e.m.}$ of the simulation analyzed in the Letter, especially in the imbalanced case. In run A, this may be caused by the fact that, due to the MHD conservation of $\sigma_{_C}$, imbalanced turbulence tends toward an equilibrium state where $\textbf{u}$ is parallel to $\textbf{B}$, an effect known as \enquote{dynamic alignment}. Consequently, since the e.m. work is proportional to $\textbf{u}\times\textbf{B}$, dynamic alignment may weaken  $W_{e.m.}$, especially at low $\omega$ as these are associated with long-lived, near-equilibrium structures. In the case of run B, the large $\beta$ implies that the kinetic pressure is much larger than the magnetic pressure. As a consequence, pressure forces may dominate over e.m. forces, especially in low frequency, near-equilibrium structures, possibly explaining why $W_{e.m.}$ is weaker at large $\tau$ in the large $\beta$ case.

\bibliographystyle{apsrev4-2}
\bibliography{CGMHD}
%\input{CGMHD.bbl}

% Main run:
% kp_int = 6.45
% kz_int = 0.79
% tnl/tA = 5.73
% drho/rho0 = 0.08
% du/cA = 0.17
% dB/B0 = 0.18

% Imbalanced run:
% kp_int = 5.13
% kz_int = 0.79
% tnl/tA = 7.21
% drho/rho0 = 0.06
% du/cA = 0.17
% dB/B0 = 0.17
% sigma_C = 0.73

% High beta run:
% kp_int = 5.79
% kz_int = 0.71
% tnl/tA = 7.75
% drho/rho0 = 0.003
% du/cA = 0.14
% dB/B0 = 0.15

%%%% 1/f? 
% We do not see it and it may be because of different reasons. One is that we do not run it for long enough, so we do not resolve small enough frequencies. Another point is that it is not well defined whether the 1/f should appear as a spatial feature (cite the expanding box paper), a temporal feature (cite Dmitruk), a mix of both, or if is something related to the fact that long time measurements sample different things with different correlation times (cite Matthaeus), rather than being a dynamical effect. So the 1/f may results from the composition of different phenomena taking place close to the Sun, rather than from turbulence alone. This is one point we will investigate in future studies. In this paper, we just want to explain the origin and role of low frequency modes that are always observed in simulations and observations of turbulence, with a clear and well defined distinction between the temporal and spatial behavior (which is not that clear in our opinion when it comes to the problem of the 1/f noise).

\end{document}